\documentclass[twocolumn,showpacs,10pt,floatfix] 
{revtex4-1}

\usepackage{amsmath}
\usepackage{amsfonts}
\usepackage{amssymb}
\usepackage{mathrsfs}
\usepackage{latexsym}
\usepackage{graphicx}
\usepackage{graphics,psfrag}
\usepackage[justification=raggedright]{subcaption}
\usepackage[font=small,justification=raggedright]{caption}
\newcommand{\nn}{\nonumber}

\begin{document}

\title{Efficiency gain and bidirectional operation of quantum engines with decoupled internal levels}

\author{Thiago R. de \surname{Oliveira}}
\email{troliveira@id.uff.br}
\author{Daniel \surname{Jonathan}}

\affiliation{Instituto de F\'{\i}sica, Universidade Federal Fluminense, Av. Gal. Milton
Tavares de Souza s/n, Gragoat\'a 24210-346, Niter\'oi, RJ, Brazil}

\begin{abstract}

We present a mechanism for efficiency increase in quantum heat engines containing internal energy levels that do not couple to the external work sink. The gain is achieved by using these levels to channel heat in a direction opposite to the one dictated by the Second Law. No quantum coherence, quantum correlations or ergotropy are required. A similar mechanism allows the engine to run `in reverse' and still produce useful work. We illustrate these ideas using a simple quantum Otto cycle in a coupled-spin system. We find this engine also exhibits other counter-intuitive phenomenology. For example, its efficiency may increase as the temperature difference between the heat baths decreases. Conversely, it may cease to operate if the hotter bath becomes too hot, or the colder bath too cold.

\end{abstract}
\maketitle

There has been a resurgence of interest in Quantum Thermodynamics \cite{Binder18},  in particular in microscopic engines operating in the quantum regime  \cite{Alicki79,Kosloff84}. 
A major goal is to understand how these quantum
engines may differ from their classical counterparts. It is already known, for example, that energy-basis coherence may boost engine power \cite{Scully11, Uzdin15}, a phenomenon that has recently been observed experimentally  \cite{Klatzow19}.  A closely related mechanism \cite{Niedenzu18,Ghosh19} is to exploit ergotropy - energy extractable by cyclic unitary rotations, without changing a system's entropy \cite{Allahverdyan04} to boost engine efficiency. This may also occur when at least one heat reservoir interacting with the engine is not in thermal equilibrium \cite{Rossnagel14,Niedenzu18,Ghosh19}, 
 in particular when it is actually a quantum measuring device 
\cite{Elouard17,Elouard18,Buffoni19}. In contrast, no firm link has been found between efficiency gains and the presence of quantum correlations between subsystems of an engine's working medium \cite{Albayrak13,He12,He12B,Huang13,Zhou12,Zhang07,Zhang08,Wang12,Huang18,Altintas14,Hewgill18}. 

\begin{figure}[t]
\begin{subfigure}[t]{0.49\columnwidth}
\includegraphics[width=\columnwidth]{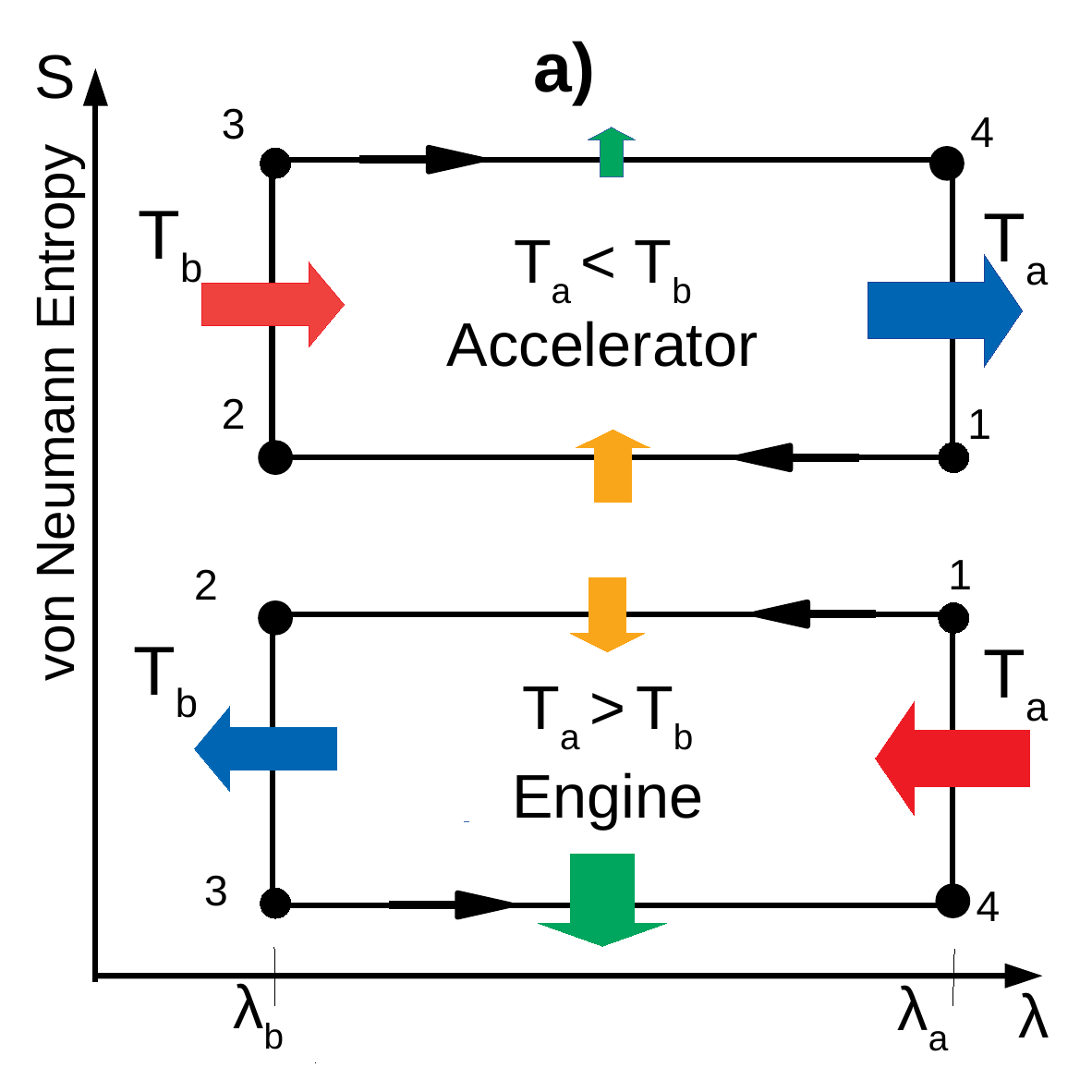}
\end{subfigure}
\begin{subfigure}[t]{0.49\columnwidth}
\includegraphics[width=\columnwidth]{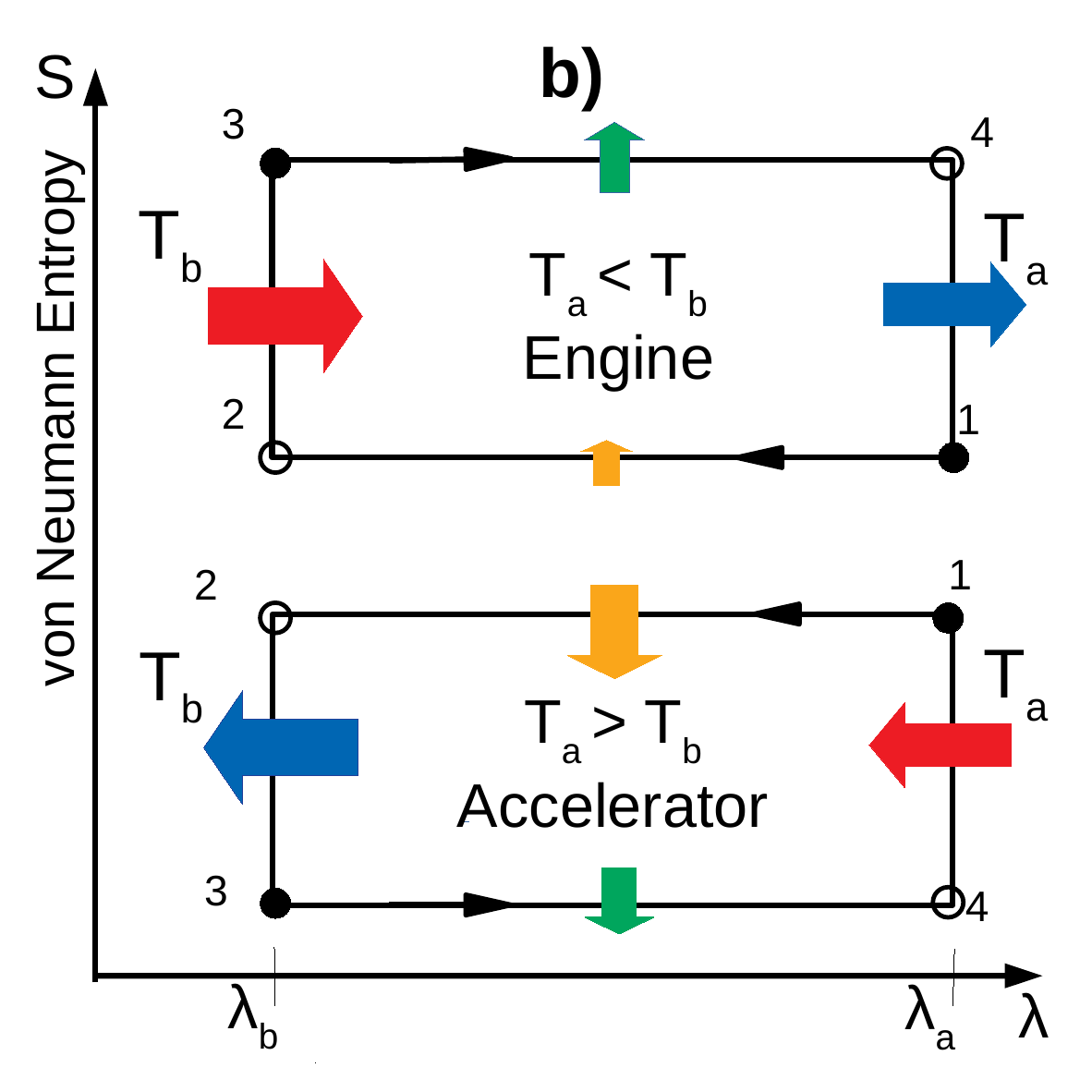}
\end{subfigure}
\caption{ (Color online)
(a) Otto cycles for a quantum system whose energy levels all follow Eq.~(\ref{eq:workinglevels}) during the adiabatic (horizontal) strokes. In this case, a heat engine must operate anti-clockwise. A clockwise cycle can represent either an accelerator \cite{Buffoni19} (depicted) or a refrigerator (not depicted).
(b) If there are also `idle' levels, independent of $\lambda$, an engine may also operate clockwise, and an accelerator anti-clockwise.  Filled (resp. empty) circles represent thermal equilibrium (resp. nonequilibrium) states. Other operation modes are possible in either cycle sense, see Fig.~\ref{Fig-Phases}.}
\label{Fig-Cycle}
\vspace{-1pt}
\end{figure}

In this article, we point out a simple alternative mechanism by which a quantum engine operating between two thermal heat baths may also achieve increased efficiency. It applies in scenarios when the working medium has discrete internal levels that do not couple to the external work sink. These levels may be made to channel heat in a reversed direction (i.e., from the cold bath to the hot), thus allowing more heat to be converted to work via the coupled levels. A similar mechanism allows the engine to produce work while running the same thermodynamic cycle in reverse (Fig.\,\ref{Fig-Cycle}). 
We also identify associated effects regarding such engines' responses to changes in the baths' temperatures. For example, they may sometimes become less efficient, or even stop working altogether, when the temperature difference increases. 

\textbf{Otto cycles} -- 
We restrict our discussion to the context of quantum Otto cycles \cite{Geva92, Kieu04, Quan07, Zhang14, Feldmann18}, although most of the concepts are more generally applicable.
These cycles use a working medium described by a Hamiltonian $H(\lambda)$, with energies $E_{n}(\lambda)$, depending on some external parameter $\lambda$. They consist of  two `isochoric' \cite{Fermi56} strokes and two (quantum) adiabatic strokes. In the former, 
$\lambda$ is fixed at values $\lambda_{a}$ or $\lambda_{b}< \lambda_{a}$ while the system equilibrates with an external thermal reservoir at respective temperatures $T_{a}$, $T_{b}$. 
The corresponding thermal states have eigenvalues  
$p_{n}^{j}~\equiv~ e^{-E_{n}^{j}/T_{j}} / Z_{j}$, 
where $E_{n}^{j}\equiv E_{n}(\lambda_{j})$  and $Z_{j} =\sum_{n} e^{-E_{n}^{j}/T_{j}}$.
In the adiabatic strokes, the system is isolated from the reservoirs while $\lambda$ is varied back or forth between $\lambda_{b}$ and $\lambda_{a}$. Supposing  that no level crossings occur, and that this variation is sufficiently slow, the populations of each level remain constant by the quantum adiabatic theorem.  For simplicity, in this article we mostly consider this idealized - but experimentally feasible \cite{Du08, Deng18, Peterson19} - regime. In some situations, however, the same evolution can be achieved for $\lambda$ driven in finite time (see e.g. the Model below).

In these cases it is straightforward to calculate the average heat and work exchanges \cite{Alicki79,Kosloff84}.  During the 
isochoric strokes $H$ is fixed, so only heat is exchanged: $Q_{a}  =\sum_{n} E_{n}^{a}\Delta p_{n};  \;\;\;  Q_{b} = \sum_{n} -E_{n}^{b}\Delta p_{n}\label{eq:Qgen}$, where $\Delta p_{n}\equiv p_{n}^{a}-p_{n}^{b}$. During adiabatic strokes entropy is fixed, and only work is exchanged. The overall extracted work is 
\begin{align}
W_{cycle} &= \sum_{n} \Delta p_{n} \left(E_{n}^{a}-E_{n}^{b}\right) \label{eq:Wgen}
\end{align}

We will find it useful below to interpret each term in these sums  as a separate energy flow, e.g. to view $q_{n}^{a} \equiv E_{n}^{a}\Delta p_{n}$ as the heat  exchanged with bath $a$ \emph{via} level $n$.

It is sometimes claimed \cite{Alicki18,Feldmann18} that the efficiency $\eta$~=~$W_{cycle}/Q_{hot}$  of these idealized Otto cycles must  equal  $\eta_0 =  1-\frac{1}{r}$, where $r = \lambda_{a}/\lambda_{b} >1$ . However, in general this only holds for working media whose energy gaps \emph{all shift in proportion to $\lambda$}, i.e., whose  energies satisfy 
\begin{align}\label{eq:workinglevels}
E_{n}(\lambda) = A(\lambda) +c_{n}\lambda,
\end{align}
for  constants $c_{n}$ and some function $A(\lambda)$ independent of $n$. Examples include a harmonic oscillator with variable frequency \cite{Kosloff17},  uncoupled spins in a variable magnetic field \cite{Geva92, Kieu04} and even some coupled spin systems \cite{Kosloff02}. 
For such systems, an adiabatic stroke maps an initial thermal state to another at a different temperature.
Thus, in these cases, the quantum notion of an adiabatic evolution coincides with the thermodynamic one \cite{Quan07}. 
 
In general, however, a quantum adiabatic stroke will drive a medium out of equilibrium \cite{Quan07}, although it will still remain in a passive \cite{Allahverdyan04},  energy-diagonal state. As we now show, this can affect $\eta$.
Consider for instance a minimal departure from the previous scenario, where now only a subset $\{E_{n}\}_{n\in \mathcal{W}}$ of the energy levels are `working', i.e. vary with $\lambda$ as in Eq.~(\ref{eq:workinglevels}), while the remainder are `idle', or independent of $\lambda$. For simplicity, we assume $A(\lambda) \equiv 0$ \footnote{Eqs.\,(\ref{eq:enhancedeff},\ref{eq:enhancedeffb}) still hold for arbitrary $A(\lambda)$, under a global energy shift such that $\lambda_{a}A(\lambda_{b}) =\lambda_{b}A(\lambda_{a})$.}.
Assuming for the moment $T_{a}>T_{b}$, so $Q_{a}=Q_{hot}>0$,
then Eq.\,(\ref{eq:Wgen}) implies:

\begin{align}
 \frac{\eta}{\eta_{0}}  = 1 -\frac{\sum_{n\notin \mathcal{W}} q_{n}^{a}} {Q_{a}}. \label{eq:enhancedeff}
\end{align}

\begin{figure}[t]
\includegraphics[width=\columnwidth]{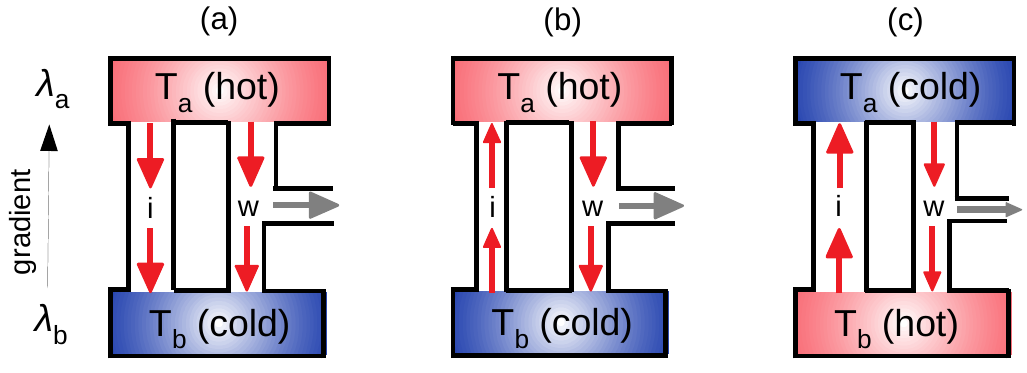}
\caption{(Color online) Energy transfer diagrams for various heat engine regimes. The left and right channels represent respectively heat flowing via idle (i) and working (w) levels. (a) An `ordinary' engine, where both channels flow from the hot bath to the cold. (b) An engine cycling in the `ordinary' sense (Fig. \ref{Fig-Cycle}(a)), but with enhanced efficiency $ \eta > \eta_{0}$. (c) A `counter-rotating' engine, as in Fig. \ref{Fig-Cycle}(b). In regimes (b,c), one channel must flow in the `wrong' sense, from cold to hot. }
\label{fig:heat-fluxes}
\end{figure}

Eq.(\ref{eq:enhancedeff}) is our first main result. It says that, for $T_{a}>T_{b}$, enhanced efficiency $\eta > \eta_{0} $ occurs if $\sum_{n\notin \mathcal{W}} q_{n}^{a}<$\,0, i.e., if on the whole idle levels channel heat \emph{into} the hot reservoir (compare Figs.~\ref{fig:heat-fluxes}(a,b)\,). 
This makes physical sense: for idle levels, $q_{n}^{a} = - q_{n}^{b}$, i.e., all the heat they absorb from one bath is deposited into the other. Thus, any heat flux from $a$ to $b$ via these levels is wasted for the purpose of generating work. 
Conversely, for a given fixed $Q_{a}$, reversing this flux allows more heat to flow via working levels, increasing $\eta$.
For idle levels with $E_{n} > 0$, such a reversal is achieved when their equilibrium population at $a$ (i.e., at higher $T$ and $\lambda$) is lower than at $b$, and vice-versa for those with $E_{n} < 0$. 
Note this need not violate the Second Law, which only 
constrains the \emph{overall} heat transfers in a cycle.

If $T_{b} > T_{a}$, the roles of hot and cold bath are exchanged, i.e.:  $Q_{b}=Q_{hot}>0$. In this case the cycle runs clockwise (Fig.~\ref{Fig-Cycle}) and ordinarily we expect it to represent either a refrigerator (where heat flows out of the colder bath and into the hotter) or an \emph{accelerator} \cite{Buffoni19} (where heat flows out of the hotter bath and, at a higher rate, into the colder).  However, now
\begin{align}
\frac{\eta}{\eta_{0}}  = r\left(\frac{\sum_{n\notin \mathcal{W}} q_{n}^{b}} {Q_{b}}-1\right) = -r \frac{\sum_{n\in \mathcal{W}} q_{n}^{b}} {Q_{b}}. \label{eq:enhancedeffb}
\end{align}

Eq.~(\ref{eq:enhancedeffb}) is our second main result. It shows that, for $T_{b} > T_{a}$, a heat engine 
requires $\sum_{n\in \mathcal{W}} q_{n}^{b}<0$, i.e. now the \emph{working} levels must channel heat in the `wrong' sense (Fig.~\ref{fig:heat-fluxes}(c)).  
We show below that such a `counter-rotating' engine is not only possible, but occurs quite generally, for example when the ground state is idle. These engines can also operate at
efficiency $\eta > \eta_0$ if the reverse heat flux is sufficiently strong ($>Q_{b}/r$).

\smallskip  

\textbf{Model} -- 
In the remainder of this article we study these behaviours, along with other associated effects, using a simple  model \cite{Zhang07,Thomas11}
of two spin-1/2 particles coupled via the Heisenberg interaction and placed in an external static magnetic field $\vec{h} = h\hat{z}$ :
\begin{equation}
H = 2J \; \vec{\sigma_1} \vec{\sigma_2} + h (\sigma_1^z + \sigma_2^z) - 2J
\label{eq:H}
\end{equation}
with $\vec \sigma_i = \{\sigma_i^x,\sigma_i^y,\sigma_i^z\}$ the Pauli matrices for site $i$ and $J\geq0$ the antiferromagnetic exchange coupling \footnote{We assume units where the magnetic moment in direction $z$  is $\mu_{z} = \pm1$, and Boltzmann's constant $k_{B} = 1$.}. This Hamiltonian is diagonal in the total spin basis. The constant $-2J$ is added so that its energies have the simple forms $-8J$, $-2h$, $0$ and $+2h$. For definiteness, we will refer to them as $E_{1,2,3,4}$, in this order, regardless of the relative magnitudes of $J$ and $h$.
We are interested in cycles where the field strength $h>0$ plays the role of $\lambda$. Note  $E_{2-4}$ satisfy Eq.~(\ref{eq:workinglevels}), with  $A(h) \equiv 0$ and $c_{2,3,4}=-1,0,+1$. However, $E_{1}$ is (nontrivially) idle if  $J> 0$. Note finally that, since the energy eigenbasis is independent of $h$, a time-dependent variation of $h$ does not change energy level populations. In other words, `adiabatic' strokes can here in fact be realised in finite time.

For this model: 
\begin{align}
Q_a &= -8J(p_1^a-p_1^b) + 2h_a(f_{b}-f_{a})\nn \\
Q_b &= -8J(p_1^b-p_1^a) + 2h_b(f_{a}-f_{b}) \nn \\ 
W_{b \rightarrow a} &=   2(h_a-h_b) f_{b}  > 0 \label{eq:work1}\\
W_{a \rightarrow b} &= - 2(h_a-h_b) f_{a}  < 0 \nn  \\
W_{cycle} &= W_{b \rightarrow a} + W_{a \rightarrow b}    = 2(h_a-h_b) (f_{b}-f_{a}), \nn
\end{align}
where
\begin{align}
f_{j}(h,J, T) \equiv p_{2}^{j} - p_{4}^{j} = \frac{2\sinh 2h_j/T_j}{1+e^{8J/T_{j}}+2\cosh[2h_{j}/T_{j}]}\label{eq:f}
\end{align} 
Note that $W_{cycle}> 0$  if and only if $f_{b} > f_{a}$. 

Although the effects we are interested in only occur for coupled spins ($J >0)$, it is useful to review what happens when $J=0$ \cite{Geva92, Kieu04, Quan07}. 
In this case
$f(h,J,T)$  reduces to
$\sinh(2h/T)/(1+\cosh(2h/T))$, which increases monotonically with $h/T$, so $f_{b} > f_{a} \iff T_a > r T_{b}$.  The same condition also implies $Q_{a} >0 , Q_{b} <0$. Thus, the cycle operates as a heat engine only if it is performed anti-clockwise on an entropy-field diagram (Fig.~\ref{Fig-Cycle}(a)) 
\footnote{Note this is reversed from the usual \emph{clockwise} representation of an ideal gas heat engine cycle on a $P-V$ or $S-V$ diagram. The physical reason is simple: as $h$ increases adiabatically, so does the gap between the ground and excited states of each spin, which shift equally in opposite directions. This leads to a net loss of energy since the ground state has larger population. However, it takes a higher temperature to maintain the same populations across a wider gap. Thus, noting also Eq. (\ref{eq:work1}), work $W_{b \rightarrow a}$ is \emph{extracted} from the system as $h$ and $T$  \emph{increase}, and vice-versa}.

Its efficiency is $\eta_0  =  1 - \frac{1}{r}$, as expected. 
Since $T_{a} > r T_{b}$, clearly $\eta_0 < 1-\frac{T_{b}}{T_{a}} \equiv \eta_{Carnot}$.

If we decrease $T_a$ so that $f_{a} > f_{b}$, then $Q_{a}, Q_{b}, W_{cycle}$ all change sign, and the cycle becomes clockwise.  Two other regimes of
operation are then possible: for $T_{b}~<~T_a < r T_b$ we have a refrigerator, and for $T_a \leq T_b$ an accelerator (Fig \ref{Fig-Cycle}(a)), since bath $b$ is now the hot bath. These behaviours mirror those of an ideal-gas Otto cycle.

\begin{figure}[t]
\begin{subfigure}{0.49\columnwidth}
\includegraphics[width=\columnwidth]{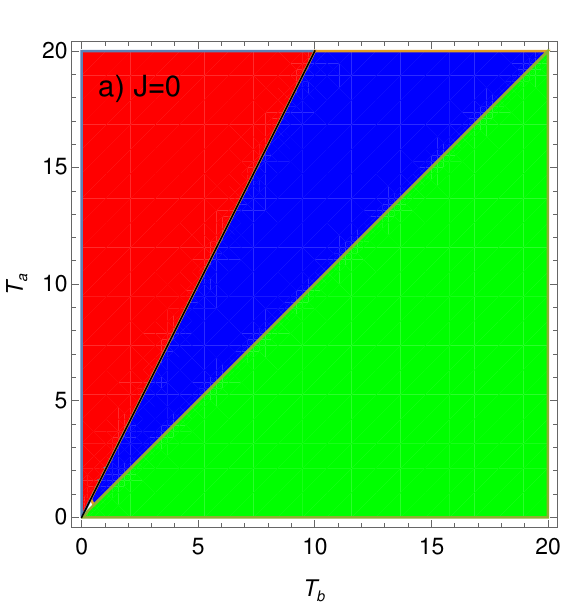}
\end{subfigure}
\begin{subfigure}{0.49\columnwidth}
\includegraphics[width=\columnwidth]{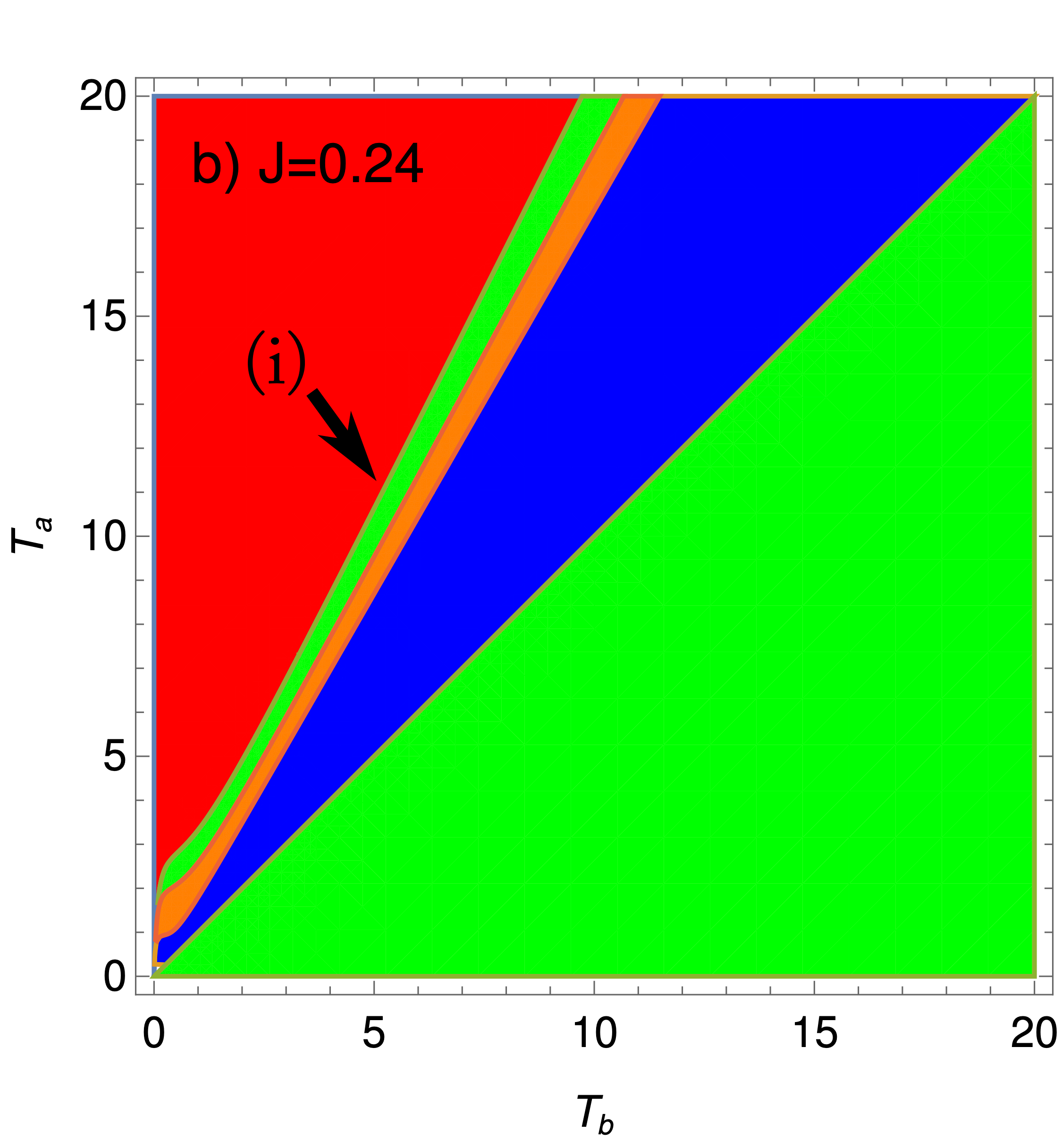}
\end{subfigure}
\begin{subfigure}{0.49\columnwidth}
\includegraphics[width=\columnwidth]{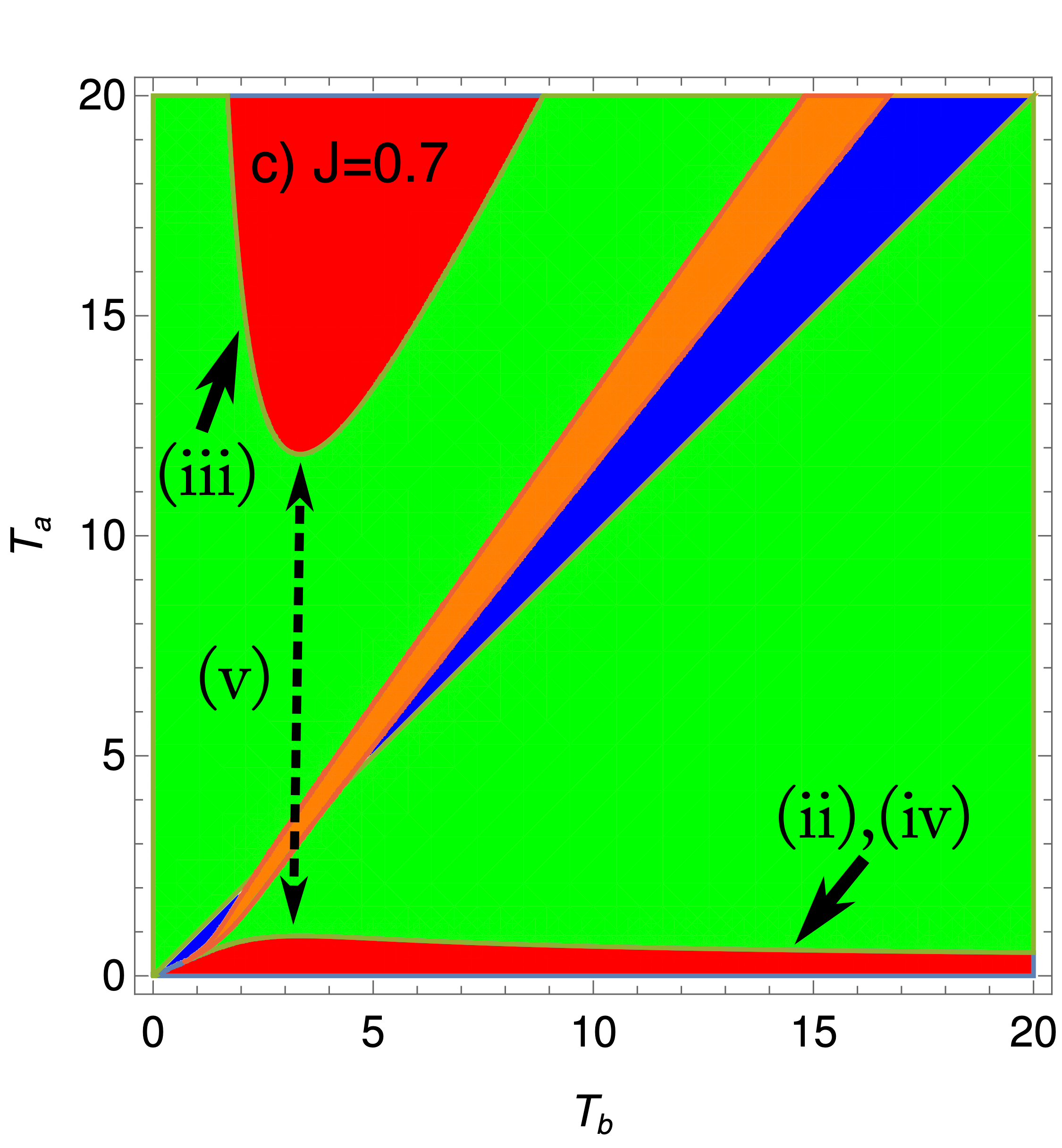}
\end{subfigure}
\begin{subfigure}{0.49\columnwidth}
\includegraphics[width=\columnwidth]{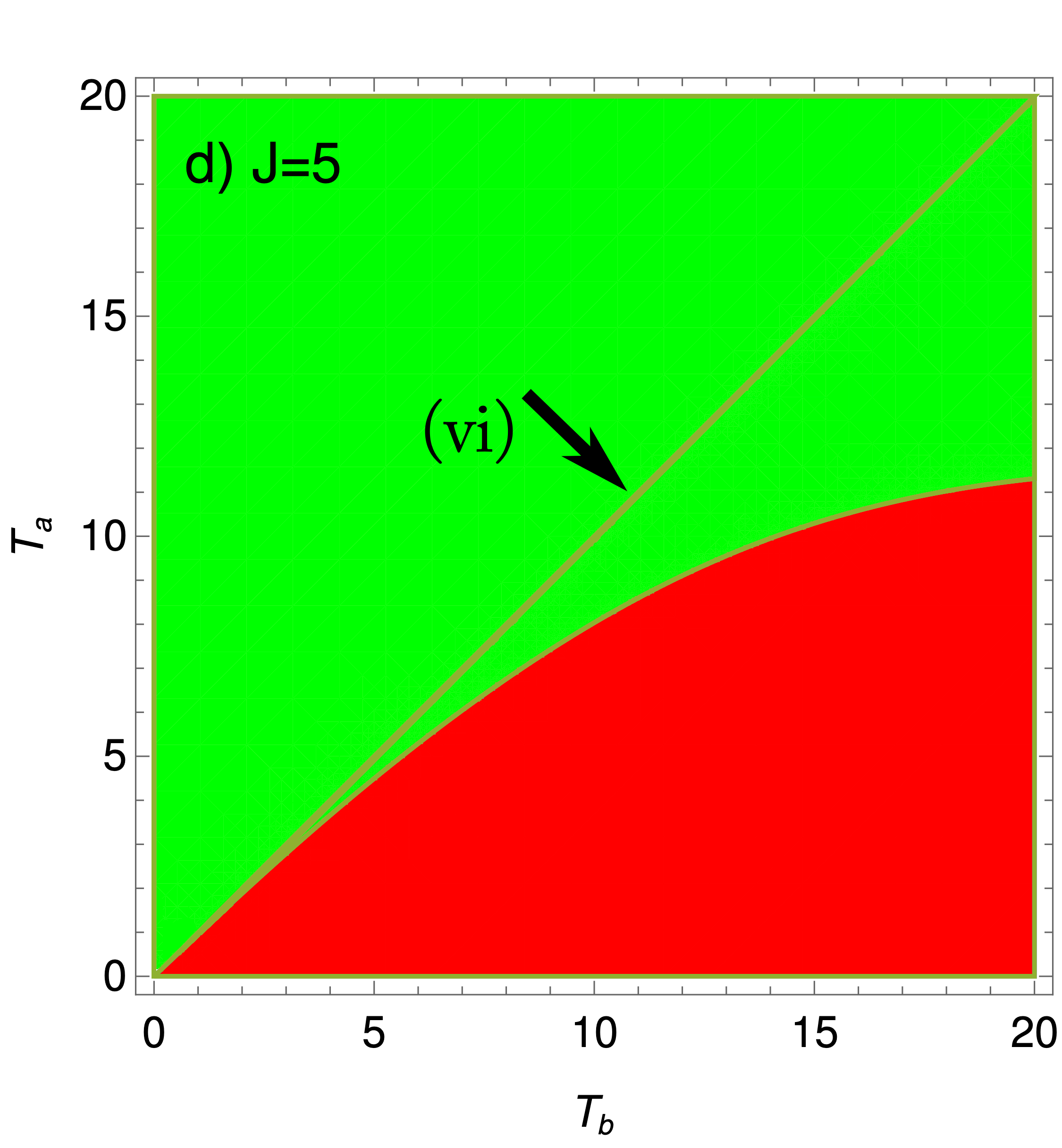}
\end{subfigure}
\caption{(Color online) Temperature ranges where Eq.\, (\ref{eq:work1}) describes a refrigerator (blue/darkest grey),  engine (red/darker grey), heater (orange/lighter grey), or accelerator (green/lightest grey), for various coupling strengths $J$. Field values are $h_a=2,h_b=1$.  $J, T, h$  are in the same (arbitrary) energy units \cite{Note2}. Several remarkable phenomena (i-vi) can be inferred (see text).}
\label{Fig-Phases}
\end{figure}

{\bf Counter-intuitive phenomenology} --
For coupled spins, the existence of the `idle' level $E_{1} = -8J < 0$ allows the two unconventional engine regimes depicted in Fig.\,\ref{fig:heat-fluxes}(b,c) to become possible. These effects are part of a wider pattern of phenomena
that run counter to (at least these authors') usual expectations for a thermal machine. To help appreciate this, 
in Fig.\,\ref{Fig-Phases} we plot, for several values of $J$, the ranges of temperatures $T_{a}, T_{b}$ for which the cycle operates in each of the four regimes allowed by Thermodynamics \cite{Buffoni19}: engine ($E$), refrigerator ($R$), accelerator ($A$) or \emph{heater} ($H$) (where work is converted into heat entering both baths).  As noted above, for $J=0$ only  the first three possibilities occur,  separated by the lines $T_{a} = r T_{b}$ and $T_{a} = T_{b}$ (Fig. \ref{Fig-Phases}(a)). As $J$ increases, however, new features appear: 

(i)~For low $J$, the $E$- and $R$-zones shrink, while
a second $A$-zone appears, and also the previously forbidden heater regime (Fig. \ref{Fig-Phases}(b)).

(ii)~For sufficiently large $J$, a second $E$-zone appears, within the region where $T_{b} > T_{a}$ (Fig. \ref{Fig-Phases}(c,d)). This is the counter-rotating (clockwise) engine regime, described in our second main result, and in Figs.\,\ref{Fig-Cycle}(b) and \ref{fig:heat-fluxes}(c). Note that, for this specific model, this phenomenon was noticed, but not fully explored, in Ref. \cite{Thomas11}. 

(iii)~`Ordinary' (anti-clockwise) engines are also still achievable at these $J$'s, within the region where $T_{a} > rT_{b}$. However, for any finite $T_{a}$, an engine becomes \emph{impossible} for sufficiently small $T_b$.
This is remarkable - we usually expect that, the lower the cold bath temperature, the \emph{easier} it is to run an engine;

(iv)~Conversely, for $T_{b} > T_{a}$ an engine becomes impossible when $T_{b}$ becomes too \emph{high}. (Notice the downward slope of the lower $E$-zone in Fig. 2(c), for large enough $T_{b}$). Thus, the hot bath can  become too hot!

(v) The two $E$-zones are separated  by a \emph{gap} in $T_{a}$, within which no heat engine is possible, regardless of $T_{b}$. 

(vi)~For larger $J$, the $H$- and $R$-zones collapse to the line $T_{a} = T_{b}$ (Fig. \ref{Fig-Phases}(d)).

\begin{figure}[t]
\begin{subfigure}[t]{0.49\columnwidth}
\includegraphics[width=\columnwidth]{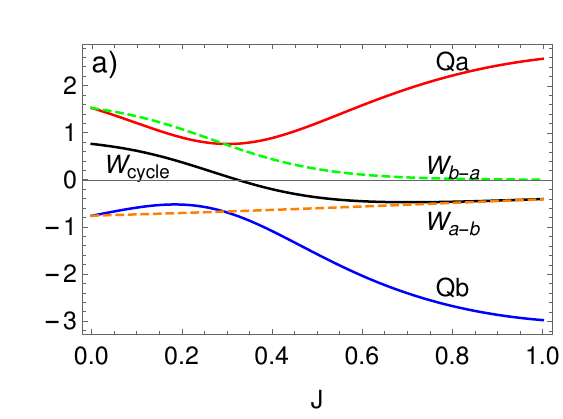}
\end{subfigure}
\begin{subfigure}[t]{0.49\columnwidth}
\includegraphics[width=\columnwidth]{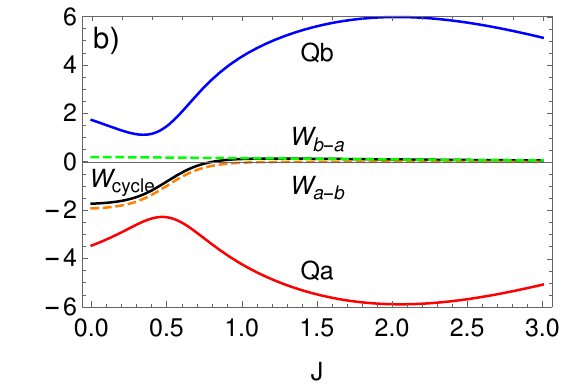}
\end{subfigure}
\caption{(Color online) Work and heat as a function of the coupling $J$ for $h_a=2,h_b=1$. (a) $T_a=5, T_b=1$. As $J$ grows, the cycle transitions from an engine to an accelerator. (b) $T_a=1$, $T_b=10$. As $J$ grows, the cycle transitions from an accelerator to a (very inefficient) engine.} 
\label{Fig-W-Q-Eng-Acc}
\end{figure}

\begin{figure}[t]
\begin{centering}
\includegraphics[width=0.49\columnwidth]{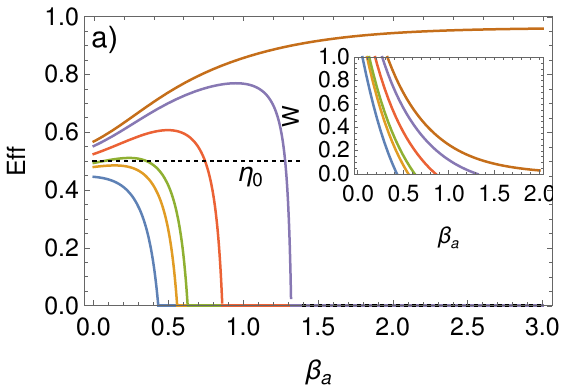}
\includegraphics[width=0.49\columnwidth]{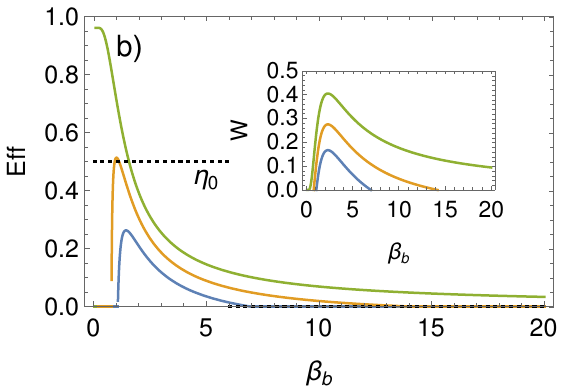}
\end{centering}
\vspace{-20pt}
\caption{(Color online) Efficiency $\eta$ for fixed $J$, $h_{b} =1,$ $h_{a} = 2$, and various temperature combinations. \newline
 (a) $\eta $ vs. $\beta_{a} \equiv T_{a}^{-1} $ for $T_b=0.2,0.1,0.08,0.05, 0.03,0.001$ (from lower to upper curves) and $J = 0.24$.  For $\beta_{a}\rightarrow 0$, $\eta$ 
surpasses the uncoupled efficiency $\eta_{0} = 0.5$
if $T_b$ is under the threshold value  $\frac{2h_{b}(h_{b} - 4J)}{\ln 3} = 0.073$ (see Appendix B).
It increases even further with $\beta_{a}$  (i.e., as the temperature difference between the baths \emph{decreases}). However, the work output $W_{cycle}$ 
 becomes exponentially small (inset). 
(b) $\eta $ vs. $\beta_{b}\equiv T_{b}^{-1} $ for $T_a = 0.04, 0.03, 0.003$ (from lower to upper curves) and $J = 0.51$. In this case $\eta > \eta_{0}$ is achievable with finite work output  (inset).}
\label{Fig-Eff-Limit}
\end{figure}

{\bf Physical Interpretation} -- This phenomenology can again be physically understood from the level structure, in particular from the fact that level $E_{1} = -8J$ is `idle'. For example:  the
condition  $J > \frac{h_a}{4}$, valid in Fig. \ref{Fig-Phases}(c,d), implies that $E_{1}$ remains the ground state throughout the cycle. In particular, it concentrates most of the system's population after thermalization with bath $a$. 
Since, however, it does not shift with $h$, then for sufficiently low $T_{a} \ll  T_{b}$  the invested work $|W_{a \rightarrow b}|$ must become smaller than the extracted work $W_{b \rightarrow a}$. This
explains the `counter-rotating' engine (property (ii)). Note also that, by Eqs.\,(\ref{eq:work1}, \ref{eq:f}), both work exchanges fall rapidly with $J$, so we might expect engines in this regime to have a very small output. While this does happen in some cases (eg Fig. \ref{Fig-W-Q-Eng-Acc}(b)), a non-vanishing work output is nevertheless still possible in the limit where $T_{a}$ is very small, see the inset in Fig. \ref{Fig-Eff-Limit}(b).

Property (iii) follows from the analogous argument in the opposite limit $T_{a} \gg T_{b}$.
In fact, it is clear this same mechanism, hence properties (ii) and (iii), apply to \emph{any} quantum `working substance' with an `idle' ground level, and for which  $W_{b \rightarrow a} $ always has the same sign.

For property (iv) note that, for high $T_{b}$, the populations $p^{b}_{j}$ all become close. Due to the symmetry of levels $\pm2h$, exponentially little work is then extracted in the expansion stroke ($W_{b \rightarrow a}\rightarrow 0$). Thus, $W_{cycle} <0$ for sufficiently high $T_{b}$. 

To understand property (i), note first that when $J=0$ the spins remain in thermal states throughout the cycle. They must therefore lose entropy when ceding heat,
and vice-versa. In particular, during a `heater'  cycle they would lose entropy in both isochoric strokes. But since there is no other source of entropy, they would then be unable to return to their initial state. Thus, a heater cycle cannot happen.  When $J \neq 0$, however, the spins leave equilibrium during the adiabatic strokes. This makes it possible, under some circumstances, for them to \emph{gain} entropy while ceding heat to a thermal bath. (If this idea sounds strange, consider a system initially in a high-energy eigenstate thermalizing in contact with a low-temperature bath). This is what allows a heater cycle.

Returning now to Eq.~(\ref{eq:work1}),
note that, for $T_{a} > T_{b}$, $W_{b \rightarrow a}$  decreases with $J$ at a faster rate than $|W_{a \rightarrow b}|$.  Consider a point just inside the $E$-zone of Fig. \ref{Fig-Phases}(a), where $W_{b \rightarrow a} \gtrsim |W_{a \rightarrow b}|$.  As $J$ grows, eventually this relation must invert, and $W_{cycle}$ becomes $<0$. By continuity, $Q_{a}>0, Q_{b}<0$ on both sides of the $W_{cycle} = 0 $ border, so this point must in fact become an accelerator (thin green region in Fig \ref{Fig-Phases}(b)). This process is illustrated in Fig. \ref{Fig-W-Q-Eng-Acc}(a). 
Other transitions, such as from refrigerator to heater to accelerator, can be similarly understood. 

In the Appendix  we analyze Fig.~\ref{Fig-Phases} in greater analytical detail.  For example, in Appendix A we prove that counter-rotating engines 
cannot exist while $J \leq~\frac{h_b}{4}$, as in Fig. \ref{Fig-Phases}(a,b), but do whenever
$J >  \frac{h_a}{4}$,
together with  the gap mentioned in property (v).
In fact, this gap is `direct'  (occurs at a specific value of $T_{b}$).  In Appendix B we derive asymptotic expressions for the inter-zone boundaries, which help us understand the shapes and disposition of the various zones, in particular property (vi).  These expressions also help us deduce asymptotic properties of the engine's efficiency (see Fig. \ref{Fig-Eff-Limit}).


{\bf Efficiency} -- Let us finally discuss engine efficiency. 
Since $n=1\notin \mathcal{W}$ then, by Eq.(\ref{eq:enhancedeff}), the cycle efficiency  when $T_a > T_b$ is
\begin{equation}\label{eq:eff1}
\frac{\eta}{\eta_0} = 1 - \frac{q_{1}^{a}}{Q_a} = 1+ \frac{8J}{Q_a}\Delta p_{1}.
\end{equation}

This expression was found, albeit without a physical interpretation, in Ref. \cite{Thomas11}. As we have noted, it means that an increase in the efficiency ($\eta > \eta_{0}$) can happen only when the heat flux $q_{1}^{a} = -8J\Delta p_{1}$ through level $1$ is negative, i.e., from the cold to the hot bath (and the increase is in fact proportional to the size of this flux). The precise conditions for this to happen depend however in a complicated way on the values of $J$ and the bath temperatures. We will not explore these details here, but we have found numerically that, for very small values of $J$, most  regimes of operation as an engine (e.g., most of the $E$-zone (red/darker grey region) in Fig.\,\ref{Fig-Phases}(b)) has $\eta>\eta_0$. However, this changes rapidly with $J$: In  Fig. \ref{Fig-Phases}(c), only a tiny region close to the origin still has enhanced efficiency, while the entire upper $E$-zone appears to have $\eta < \eta_{0}$. 


We can prove analytically, however, another counter-intuitive phenomenon involving the efficiency, illustrated in Fig. \ref{Fig-Eff-Limit}. In Fig. \ref{Fig-Eff-Limit}(a) we plot $\eta$ as a function of $\beta_{a}\equiv T_{a}^{-1} $, for various fixed values of $T_{b}$ and a fixed $J \lesssim \frac{h_{b}}{4}$. Note first that, as $\beta_{a}\rightarrow 0$, $\eta$ tends to a value above $\eta_{0}$ (dashed line) only when $T_{b}$ is under a certain threshold. Indeed, in Appendix B we prove that the threshold value is $T_{b}^{0} \equiv \frac{2h_{b}(h_{b} - 4J)}{\ln 3}$. Moreover, we also prove that in these cases the efficiency \emph{increases further} with $\beta_{a}$ (ie, as  $T_{a}$ \emph{decreases}), as can be seen in the figure \footnote{It is worth noting that $\eta$ is always upper bounded by $ \frac{\eta_{0}}{1 - 4J/ h_{a}} < \eta_{Carnot}$  \cite{Thomas11}}. This behaviour is contrary to the customary wisdom that a heat engine should become \emph{less} efficient as one decreases the temperature difference across which it operates. Physically what happens is that, as $T_{a}$ falls, both the overall absorbed heat $Q_{a}$ and the counter-propagating heat flux through level $1$ reduce in magnitude, however the latter does so at a slower rate. By Eq.~(\ref{eq:eff1}), this increases the efficiency. However, in this regime the extracted work $W_{cycle}$ also falls to 0 exponentially with $\beta_{a}$ (see inset). 

Another regime exists, however, where a similar effect happens but with $W_{cycle}$ still remaining finite. Consider the counter-rotating engine regime when $T_a \ll T_b$. In this case, as mentioned above in property (iv),  for any fixed $T_{a}$ there is a finite $T_{b}$ above which no engine is possible. It follows that, as $\beta_{b}\equiv T_{b}^{-1} $ is increased from zero, $\eta$ will at a certain point change from zero to positive and continue to increase, at least for a while. This is verified in Fig. \ref{Fig-Eff-Limit}(b). As can be seen in the inset, in these cases $W_{cycle}$ can indeed remain finite for $\eta > \eta_{0}$.

\textsl{}

{\bf Discussion} --  We have presented an alternative mechanism to achieve greater efficiency in quantum thermal machines operating via the Otto cycle. 
It applies when at least one of the system's energy  levels does not couple to the external work source.
Moreover, this mechanism also leads to counter-intuitive phenomena such as the operation of an engine in both directions of the same cycle, an efficiency increase when the bath temperature difference decreases and situations where the engine can cease to be possible if the hot bath becomes too hot, or the cold bath too cold, or when one of the baths'  temperatures lies within a specific range.
We have also given an interpretation in terms of separate heat transfers from one bath to another via each system level, showing that some of these effects occur when at least one of these fluxes is in a direction contrary to that of the overall heat flow. Finally, we have illustrated these phenomena in a simple physical model of two spins-1/2 coupled via the Heisenberg interaction.

We emphasize that, unlike in Refs. \cite{Uzdin15,Scully11,Klatzow19,Niedenzu18,Ghosh19}, neither energy-basis coherence nor ergotropy has been exploited here - the system remains always in a passive \cite{Allahverdyan04}, energy-diagonal state. Moreover, all results for our spin model would still hold if its eigenvectors were all product states - in which case the spins would remain unentangled throughout the cycle. Indeed,  they also hold for a three-level (qutrit) model with energies $-8J, \pm2h$ \cite{TeseMaron20}. This proves the presence of entanglement is not relevant for efficiency gains, confirming indications by previous studies \cite{Albayrak13,He12,He12B,Huang13,Zhou12,Zhang07,Zhang08,Wang12,Huang18}.
The only quantum aspect of our engine that seems relevant to the phenomena we have described is its discrete energy spectrum \cite{Campisi16,Gelbwaser-Klimovsky18}. Combined with the unitary isoentropic strokes, this allows the working substance to leave thermal equilibrium, which appears to be a key requisite. Note the same could occur in a fully classical stochastic model with an effective discrete spectrum and probability-preserving strokes. However, we could not locate any description in the literature of such phenomena occurring in a classical engine.

Although we have restricted our analysis to specific, idealised scenarios, we expect the phenomenology will generalise. In Appendix C we briefly discuss three possible generalisations, none of which feature `idle' levels. Preliminary findings \cite{Oliveira21} also indicate most results extend to cycles where nonadiabatic time-dependent strokes do couple different energy eigenstates, resulting in quantum friction \cite{Feldmann18}.

\begin{acknowledgments}

This  work is supported by the Brazilian National Institute for Science and  Technology of Quantum Information (INCT-IQ), and by the Air Force Office of Scientific Research under award number FA9550-19-1-0361.

\end{acknowledgments}

\appendix

\section{Analysis of the Coupled Spin Otto Cycle}

\subsection{Conditions for engines with $T_{b} > T_{a}$}

As we have noted, one of the remarkable features of this coupled spin system is that it is able to operate as a heat engine in either sense (clockwise or anti-clockwise) of the Otto cycle, as depicted in an $S-h$ diagram. However, while a `normal', anti-clockwise engine cycle (running with $T_{a} > T_{b}$) can exist for any value $J\geq 0$ of the coupling constant, a clockwise cycle (running with $T_{b} >  T_{a}$) can only operate as an engine if the coupling is sufficiently strong. More specifically: \medskip

\textbf{Result 1}:  Suppose $T_{b} > T_{a}$. (i) A cycle producing positive work $W_{cycle}>0$ is impossible if $0\leq  J \leq h_{b}/4$. (ii) Conversely, if $J > h_{a}/4$, a cycle with $W_{cycle}>0$ is possible  for all $T_{b}$ and sufficiently small $T_{a} $. \medskip

This result is illustrated in the fact that there is no red/darker grey (E) zone in the bottom right half of Figs. \ref{Fig-Phases}(a,b), but one does appear in Figs. \ref{Fig-Phases}(c,d). Moreover, this bottom $E$-zone runs along the entire $T_{b}$-axis. \smallskip

\begin{figure}[t]

\includegraphics[width=\columnwidth]{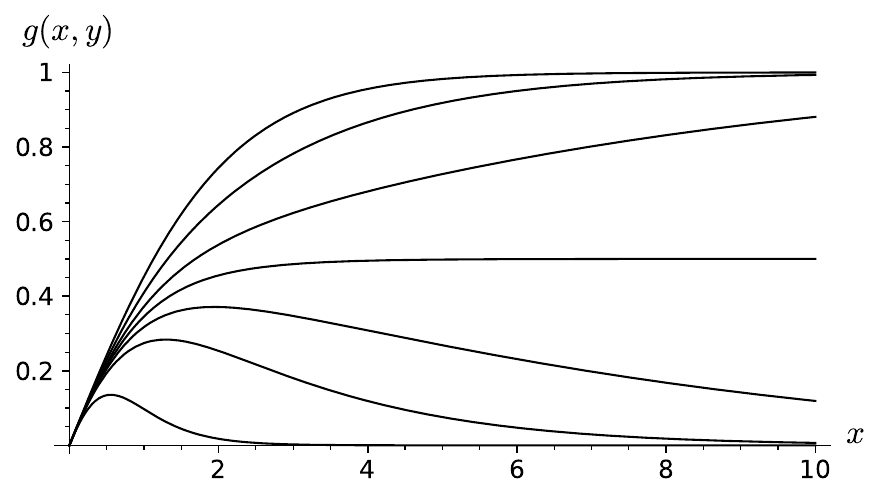}
\caption{Work function in the generic form $g(x,y)$ of Eq. (\ref{eq:g}), shown as a function of $x$ for various fixed values of $y$. \newline From top to bottom: $y = 0.1, \,0.5, \,0.8,\, 1, \,1.2, \,1.5, \, 3$. 
\newline For $y < 1$, the function increases monotonically towards 1 as $x\rightarrow \infty$. For $y = 1$, it tends asymptotically to $1/2$. For $y >1$, $g(x,y)$ is single-peaked and tends to $0$ as $x\rightarrow \infty$. }  
\label{Fig-g}
\end{figure}

\emph{Proof of Result 1}:  As shown in Eq. (\ref{eq:f}), the amount of work exchanged in each adiabatic stroke is proportional to the `work function'
\begin{equation} \label{eq:fagain}
f_{j}(h, J, T) = \frac{2 \sinh\left(2h_{j} / T_{j} \right)}{1 + 2 \cosh\left(2h_{j} / T_{j} \right) + e^{8J / T_{j}}},
\end{equation}
where $j \in\{a,b\}$ denotes the starting point of the stroke.
It is useful to rewrite this in the form 
\begin{equation}\label{eq:fg}
f_{j}(h, J, T) = g\left(\frac{2h_{j}}{ T_{j}}, \frac{4J}{h_{j}}\right),
\end{equation}
where
\begin{align}
g(x,y) = \frac{e^{x} - e^{-x}}{1 + e^{x} + e^{-x}  + e^{xy}}. \label{eq:g}
\end{align}
This function, shown in Fig. \ref{Fig-g}, has the following easily checked properties  
for $x,y \geq 0:$
\begin{align}
g(x,y) &\geq 0 \label{eq:gpositive}; \;\;\;g(x,y) =0 \iff x = 0  \\
\lim_{x\rightarrow \infty} g(x,y) & = 
\begin{cases}
0, & y > 1; \\
1/2, & y = 1; \\
1, & 0\leq y < 1. \\
\end{cases}\label{eq:limitg}\\
\frac{\partial g(x,y)}{\partial x} &> 0 \text{ for $0 \leq y \leq 1$.} \label{eq:monotx}\\
\frac{\partial g(x,y)}{\partial x} &= 0 \text{ at a single value of $x$, for $y > 1$} \label{eq:maximumg}\\
\frac{\partial g(x,y)}{\partial y} & < 0.  \label{eq:monoty} \\
g(rx, y) &> g(x,ry) \text{ for $r > 1$.} \label{eq:gscaling}
\end{align}

Suppose then that $T_{b} > T_{a}$, and $0  \leq J \leq \frac{h_{b}}{4}$, i.e., $ J = s \frac{h_{b}}{4}$, for some $0 \leq  s \leq 1$. In this case, 
\begin{align}
f_{b} - f_{a} =  g \left(\frac{2h_{b}}{ T_{b}}, s \right) - g \left(\frac{2h_{a}}{ T_{a}}, \frac{s}{r}\right) <0.
\end{align}
where we used Eq. (\ref{eq:gscaling}) and the fact that  $r=\frac{h_{a}}{h_{b}}>1$.\footnote{This conclusion also follows from Eqs. (\ref{eq:monotx}) and (\ref{eq:monoty}). } By Eq. (\ref{eq:work1}), this means $W_{cycle} <   0$, so  no engine is possible for $J$ in this range.

On the other hand, suppose $J  = s \frac{h_{a}}{4}$, for some $s > 1$. In this case
\begin{align} \label{eq:fb-fa}
f_{b} - f_{a} = g \left(\frac{2h_{b}}{ T_{b}}, r s \right) - g \left(\frac{2h_{a}}{ T_{a}}, s \right). 
\end{align}
By Eq. (\ref{eq:limitg}), $\lim_{T_{a}\rightarrow 0} f_{a} = 0$. Thus, for any fixed $T_{b}$ we must have $f_{b} - f_{a}  > 0 \Leftrightarrow W_{cycle} >  0$ when $T_{a}$ is sufficiently small $\square$\medskip

\subsection{Temperature gap}

Another striking feature of this model, visible in Fig.~\ref{Fig-Phases}(c), is the appearance of a \emph{temperature gap} for $T_{a}$, that is, a range of values $T_{a} \in (T_{a1}, T_{a2})$ for which the cycle cannot operate at all as an engine - whatever the value of $T_{b}$. For $T_{a}$ within this range, the work extracted during the expansion stroke of the cycle is always less than the one invested in the compression stroke.

We will now prove that such a gap always exists if  the coupling $J$ is sufficiently high. In fact, borrowing the terminology of solid-state physics, it is always a `direct gap', in the sense that these two extremes of the gap range occur for the same value of $T_{b}$. Graphically, the lowest point of the `upper E-zone'  is always directly above the highest point of the `lower E-zone'.

\medskip

\textbf{Result 2}:  For $J > h_{a}/4$,  there exists a range $(T_{a1}, T_{a2})$ of values of $T_{a}$  for which $W_{cycle} < 0$,  for any value of $T_{b}$. 
Furthermore: given $h_{a}, h_{b}, J$, there exists a temperature $T_{b0}$ such that
\begin{itemize}
\item[(i)] $T_{b0}$ is the value of $T_{b}$ that maximizes the work $W_{b\rightarrow a}$ extracted during the expansion stroke,
\item[(ii)] $T_{b0} \in(T_{a1},T_{a2})$, and 
\item[(iii)] $W_{cycle}(T_{a},T_{b}) = 0$ for $(T_{a}, T_{b}) = (T_{a2} ,T_{b0})$ and $(T_{a}, T_{b}) = (T_{a1} , T_{b0})$. In other words, the gap in $T_{a}$ is `direct' and occurs at $T_{b}= T_{b0}$
\end{itemize}
\medskip

\emph{Proof of Result 2}: We will base our analysis purely on the general mathematical properties (\ref{eq:gpositive})-(\ref{eq:gscaling}) of $g(x,y)$.
Note first that the borders of the $E$-zones are  defined by the condition  $W_{cycle} = 0$. For the case $J > h_{a}/4$, Eq.~(\ref{eq:fb-fa}) shows this condition is equivalent to
\begin{equation} \label{eq:Ezones}
g\left(\frac{2h_{a}}{ T_{a}}, s \right) = g\left(\frac{2h_{b}}{ T_{b}}, r s\right)
\end{equation}
where $r,s >1$.  
Since we are assuming fixed $h_{a}, h_{b}, J$ here, it is helpful   to consider each side of this equation as a function only of $T$, e.g., to redefine
\begin{equation} \label{eq:G}
g\left(\frac{2h_{a}}{ T_{a}}, s \right) \equiv G(T_{a}); \;\;
g\left(\frac{2h_{b}}{ T_{b}}, rs \right) \equiv \overline{G}(T_{b}).
\end{equation} 
\begin{figure}[t]
\includegraphics[width=\columnwidth]{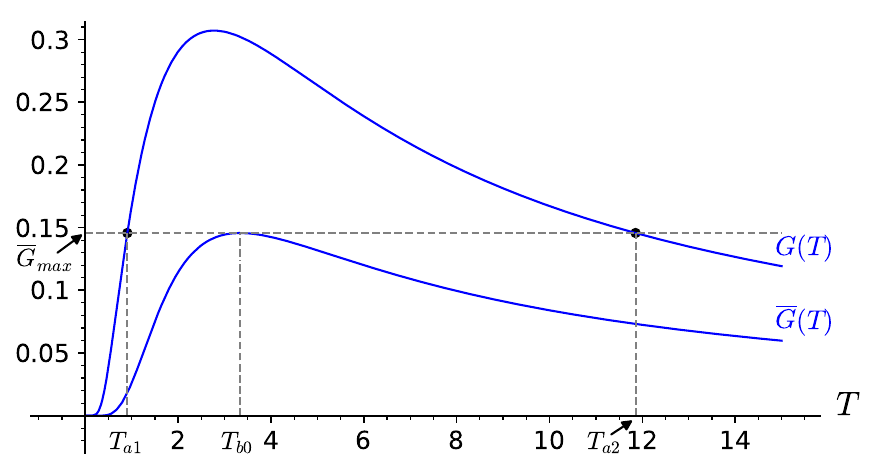} \caption{(Color online) Functions $G(T)$ and $\overline{G}(T)$ defined in Eq. (\ref{eq:G}). The highlighted dots show the two points where $G(T)$ reaches the maximum value $\overline{G}_{\max}$ of $\overline{G}(T)$. The interval  $T_{a1}<T< T_{a2}$ where $G(T) >  \overline{G}_{max}$ corresponds to the  gap in $T_{a}$ where no engine can exist in Fig.~\ref{Fig-Phases}(c). The  temperature $ T_{b0}$ which maximizes $\overline{G}(T)$ is the point in $T_{b}$ where the gap occurs. The parameters $h_{a} = 2$, $h_{b}= 1$, $J=0.7$ are the same here as in that figure. For these values, $T_{a1} = 0.91,\, T_{b0} = 3.34, \, T_{a2} =11.85$.} 
\label{Fig-Gap}
\end{figure}
Let us now analyze the temperature ranges for which Eq. (\ref{eq:Ezones}) may or may not have solutions. Since $r,s >1$, Eqs. (\ref{eq:gpositive}),(\ref{eq:limitg}),(\ref{eq:maximumg}) imply that $G(T)$ and $\overline{G}(T)$ are single-peaked in $T$, and tend to $0$ for $T \rightarrow 0, \infty$ (see Fig. \ref{Fig-Gap} above). Let $ T_{b0}$ be the temperature  where $\overline{G}(T_{b0})$ achieves its peak  $\overline{G}_{\max} \equiv \max_{T}\overline{G}(T)$. Comparing the definitions of $\overline{G}$ in Eq. (\ref{eq:G}) and $W_{b\rightarrow a}$ in Eq. (\ref{eq:work1}), using also Eq. (\ref{eq:fg}), we see this is the same as property (i). 

Now: Eq. (\ref{eq:monoty}) implies that 
\begin{align}\label{eq:Gmax}
\overline{G}_{\max} & = g \left(\frac{2h_{b}}{ T_{b0}}, r s \right) <  g \left(\frac{2h_{b}}{ T_{b0}}, s \right)
 = G(rT_{b0}) \leq G_{\max}
\end{align}
where $G_{\max} \equiv \max_{T} G(T)$.  Since $G(T)$ is single-peaked, it follows that, given any $T_{b}>0$,  Eq. (\ref{eq:Ezones}) will always be satisfied for exactly two values of $T_{a}$, that we will call $T_{a+}(T_{b}) > T_{a-}(T_{b})$. (These solutions describe the upper and lower $E$-zone borderlines in Fig. \ref{Fig-Phases}(c). 
In particular, if we define 
\[
T_{a1} \equiv T_{a-}(T_{b0}); \;T_{a2} \equiv T_{a+}(T_{b0}),
\]
then $G(T_{a1,2}) = \overline{G}_{\max} < G_{\max}$. Since $G(T)$ is single-peaked, $G(T_{a}) > \overline{G}_{\max}$ $\iff$ $T_{a}~\in~(T_{a1},T_{a2})$. But this implies there is no solution for Eq. (\ref{eq:Ezones}) for  $T_{a}$ in this interval. More specifically, within this gap   
\[
W_{cycle} \propto (\overline{G}(T_{b}) - G(T_{a}) ) < (\overline{G}_{\max} - G(T_{a}) ) < 0, \;\;\forall T_{b}.
\]
Conversely, for any $T_{a} > T_{a2}$ or $< T_{a1}$, an analogous argument shows there will be two values of $T_{b}$ that solve Eq. (\ref{eq:Ezones}). However, since $\overline{G}$ is single-peaked, for $T_{a} =T_{a2}$ or $T_{a} = T_{a1}$ the only solution is $T_{b} = T_{b0}$. Thus we obtain property (iii).

We still need to prove property (ii). For the upper bound: by Eq. (\ref{eq:Gmax}), $G(rT_{b0}) > G(T_{a2}) = G(T_{a1})$. Since $G$ is single-peaked, and by definition $T_{a2}> T_{a1}$, this implies $T_{a2} > rT_{b0} > T_{b0}$.

For the lower bound:  Eq. (\ref{eq:gscaling}) implies that, in fact, 
\[
G(T) > \overline{G}(T),\;\;\; \forall T>0.
\]
as can be seen in Fig. \ref{Fig-Gap}. But then $G(T_{b0}) > \overline{G}(T_{b0}) = G(T_{a1}) = G(T_{a2})$. Since $G$ is single-peaked, this means that $T_{a1} < T_{b0} < T_{a2}$. Note that we reobtain also the upper bound in this way. However, as seen above, the latter is independent of property (\ref{eq:gscaling}) $\square$ \medskip

In conclusion: the temperature gap in $T_{a}$ appears due to the fact that, for large enough coupling $J$, the work function $f(h, J, T)$ becomes single-peaked in  $h/T$. Physically, this occurs because, for large $J$, the ground state of this model becomes the level $-8J$, which does not contribute to the work exchanges in the adiabatic strokes, but which contains more and more of the system's population as $T \rightarrow 0$. The appearance of this gap can therefore be argued to be a quantum effect, since such a population accumulation in a single discrete state does not happen in classical systems.

It is worth emphasizing that, in the demonstration of Result 2, we only made use of properties (\ref{eq:gpositive})-(\ref{eq:gscaling}) of the work function, and not of the specific form of $f(h, J, T)$.  Since these are quite generic,  we can expect that other, more complex models with similar characteristics will also exhibit a temperature gap where no engine operation is possible. An interesting open problem is to investigate more carefully to what extent properties such as Eqs. (\ref{eq:gpositive})-(\ref{eq:gscaling})  are mandated by general constraints such as the Second Law, irrespective of the details of  the Hamiltonian. For example: note that, in the demonstration above, the only place where property (\ref{eq:gscaling}) was required was to prove $T_{a1}< T_{b0}$. In the absence of this property, a direct gap would still exist, but it might be entirely within the region where $T_{a} > T_{b}$. This would, however, imply an engine existing for $T_{a} = T_{b}$, which is forbidden by the Second Law, for we would be extracting positive work from a single thermal reservoir. We therefore conclude that, if a system satisfies properties (\ref{eq:gpositive})-(\ref{eq:monoty}), the Second Law \emph{requires} it to also satisfy property (\ref{eq:gscaling}).

\medskip

\section{Asymptotic Analysis}

\begin{figure}
\includegraphics[width=\columnwidth]{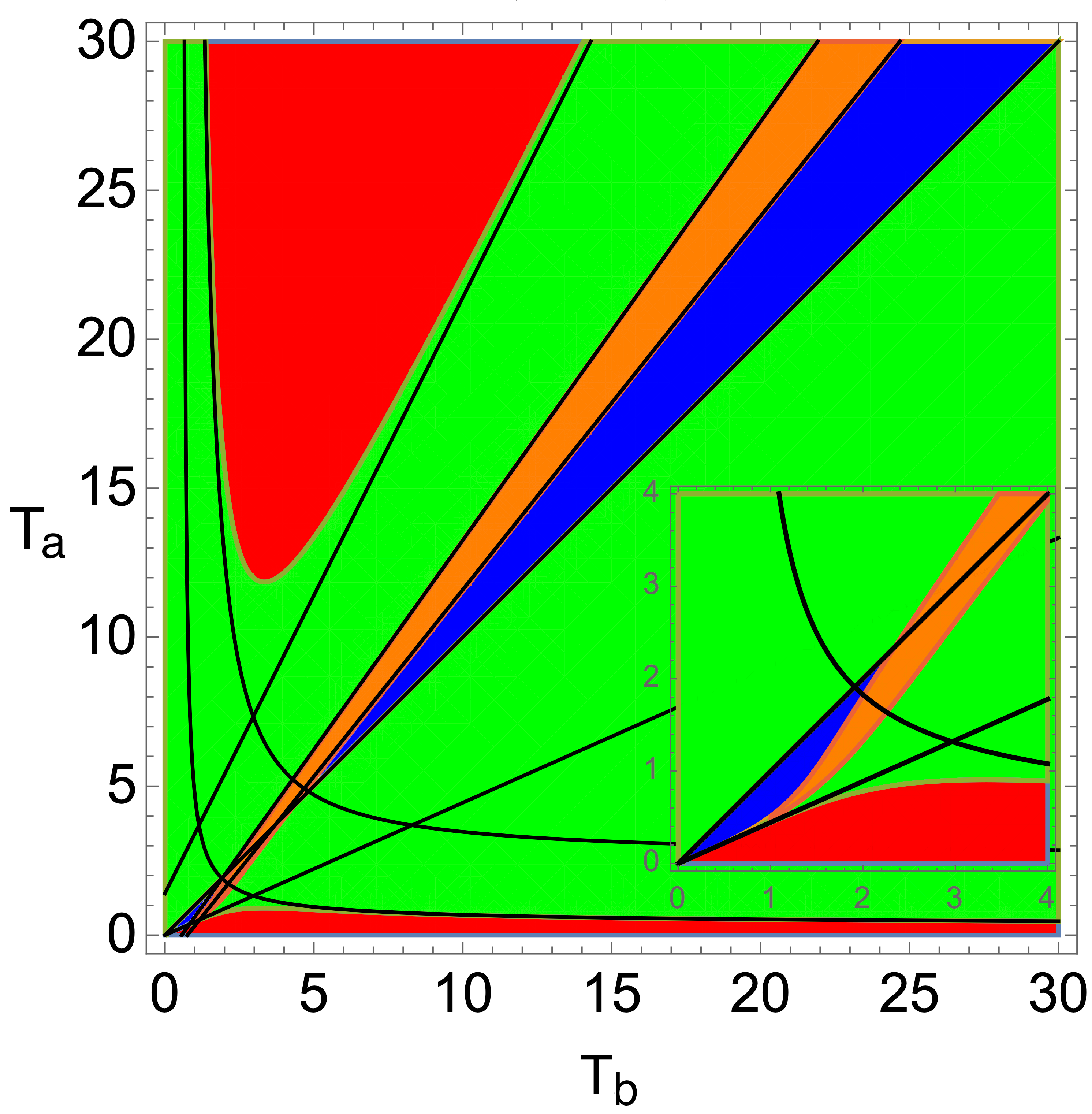}
\caption{(Color online) Asymptotic approximations for the borders of the various operation zones of the system. The system parameters are $h_{a}=2,$ $h_{b}=1,$ $J=0.7$ (same as in Fig. \ref{Fig-Phases}(c) ).  As before, the the refrigerator ($R$) zones are in blue/darkest grey, the engine ($E$) zones in red/darker grey,  the heater ($H$) in orange/lighter grey and the accelerator ($A$) in green/lightest grey. The black lines are our asymptotic approximations to the various inter-zone boundaries. For the upper $E$-zone, these are given by Eq. (\ref{eq:upperE-asymp2}) (left side) and Eq. (\ref{eq:upperE-asymp1}) (right side). For the lower $E$-zone, they are given by Eq. (\ref{eq:lowerE-asymp1}) (left side) and Eq. (\ref{eq:lowerE-asymp2}) (right side). For the \emph{A-H}, \emph{H-R} and \emph{R-A} borders, they are given respectively by Eqs. (\ref{eq:AH-asymp}),  (\ref{eq:HR-asymp})  and (\ref{eq:AR}). The inset shows a closer view of the bottom left corner, omitting Eqs. (\ref{eq:AH-asymp}) and  (\ref{eq:HR-asymp}) for clarity. Note in particular the second $R$-zone, corresponding to a `counter-rotating' refrigerator ($T_{b} > T_{a}$)}
\label{Fig-Asympt}
\end{figure}

In this section we study how the system behaves when 
the temperatures $T_{a}, T_{b}$,  become relatively `large' or `small' in comparison to each other or to
other parameters. We use the resulting asymptotic formulae to undertand the boundaries between the various zones appearing in the diagrams of Fig. \ref{Fig-Phases} (see Fig. \ref{Fig-Asympt} for greater detail). Another application is to study the asymptotic behaviour of the heat engine efficiency $\eta$ in the limit of large $T_{a}$. In particular, we demonstrate that, in this limit, $\eta$ can increase as $T_{a}$ decreases, for fixed $T_{b}$.

Our analysis is based on  the following simple approximations: 
\begin{itemize}
\item 
For  `large' temperatures, satisfying  $2h_{j}/T_{j} \sim 8J/T_{j} \sim \varepsilon \ll 1$, we can express  the work function in Eq. (\ref{eq:fagain}), and more generally the probabilities $p_{1}^{j}$ appearing in the text, as power series in $\varepsilon$:
\begin{align}
f_{j}(h, J, T) & = \frac{h_{j}}{T_{j}} - \frac{2h_{j}J}{T_{j}^{2}} + O(\varepsilon^{3}) \label{eq:f-expansion}\\
p_{1}^{j} & = \frac{1}{4}\left[1 + \frac{6J}{T_{j}} + \frac{12 J^{2}-h_{j}^{2}}{T_{j}^{2}} \right] + O(\varepsilon^{3})  \label{eq:p-expansion}
\end{align}
\item For `small' temperatures $T_{j} \ll 8J , 2h_{j}$, the appropriate approximations are
\begin{align}\label{eq:expapprox}
f_{j}(h, J, T) & \simeq \frac{1}{1+\exp\left(\frac{8J - 2h_{j}}{T_{j}}\right)}\\
p_{1}^{j}(h, J, T) & \simeq \frac{1}{1+\exp\left(\frac{2h_{j}-8J}{T_{j}}\right)}. \label{eq:expapprox2}
\end{align}
In other words, in this limit these functions behave as (complementary) step functions in $J$, with a sharp cutoff for $J > h_{j}/4$.
\end{itemize}

\subsection{Asymptotic Zone Boundaries}

Our first application is to obtain asymptotic expressions for the boundaries between the various operation regimes of the system. Figure \ref{Fig-Asympt} above illustrates the result of these calculations, which are detailed below. The figure is an expanded version of Fig. \ref{Fig-Phases}(c), with added lines corresponding to our asymptotes. We can see that agreement with the exact zone boundaries is generally extremely good, even for temperatures well within the depicted range.

\subsubsection{Boundary of the upper $E$-zone} 

The boundaries of the $E$-zones are defined by the condition $W_{cycle} =0$, which is equivalent to $f_{a} = f_{b}$ by Eq.~(\ref{eq:work1}). For the `upper'  ($T_{a} > T_{b}$) $E$-zone, we can solve this equation for $T_{a}$ in the limits where (i) $T_{a}$ and $T_{b}$ are both `large', or (ii) $T_{a}$ is `large' and $T_{b}$ is `small'.   \medskip

(i) Suppose $T_{a}\gtrsim T_{b} \gg 2J, 2h_{a}, 2h_{b}$.  In this case it is possible to find $T_{a}$  by an iterative (perturbative) method. Using Eq. (\ref{eq:f-expansion}), we solve the equation $f_{a} = f_{b}$ to $O(\varepsilon)$, then substitute the solution into the equation to $O(\varepsilon^{2})$, and so forth.

In this manner, it is possible in principle to obtain a complete        series expression for the boundary line, valid in the limit of large $T_{b}$. The solution correct up to $O(\varepsilon)$, which is sufficient for our purposes, is
\begin{align}\label{eq:upperE-asymp1}
 T_{a} & = rT_{b} + 2J (r-1) 
\end{align}
Note that when $J=0$ (no coupling), we recover asymptotically the exact $E$-zone boundary (Fig. \ref{Fig-Phases}(a)). For $J > 0$, Eq. (\ref{eq:upperE-asymp1}) shows that the boundary remains asymptotically a straight line, with the same inclination as previously, but shifted upwards by an amount proportional to $J$. Physically this means that, for a given `large' temperature $T_{b}$,  the temperature $T_{a}$ required to run an engine increases with the coupling strength. This can indeed be seen in Figs. \ref{Fig-Phases}(b,c). In Fig. \ref{Fig-Phases}(d) the upper $E$-zone exists but is not visible - it has shifted up completely out of the depicted area, since $J$ is so large. \medskip

(ii) For $T_{b} \ll 8J,  2h_{b}, 2h_{a} \ll T_{a}$, the boundary condition, correct to $O(\varepsilon) \sim h_{a}/T_{a} \sim J/T_{a}$, becomes 
\begin{align} \label{eq:upperEleft}
\frac{h_{a}}{T_{a}}\left(1-\frac{2J}{T_{a}}\right) & = \left(1+\exp\left(\frac{8J - 2h_{b}}{T_{b}}\right)\right)^{-1} 
\end{align}
Inverting both sides and using the fact that, to same order in $\varepsilon$,  $\frac{1}{\left(1-\frac{2J}{T_{a}}\right)}  = 1 +\frac{2J}{T_{a}}$, we obtain
\begin{align}\label{eq:upperE-asymp2}
 T_{a} & =  -2J + h_{a}\left(1+ \exp\left(\frac{8J - 2h_{b}}{T_{b}}\right)\right)\
\end{align}
When $J > h_{a}/4 \; (> h_{b}/4)$, Eq. (\ref{eq:upperE-asymp2}) implies that the minimum temperature $T_{a}$ necessary for an engine to be possible \emph{increases exponentially} as $T_{b}\rightarrow 0$. This remarkable property, which can be clearly seen in Fig \ref{Fig-Asympt} above, can be physically understood as follows: recall that $J > h_{a}/4$ implies that the ground state is the level $-8J$, which does not take part in work exchanges. After the system thermalises with the very cold bath at $T=T_{b}$, the population of all other levels will be exponentially small, so the same will be true for the amount of work extracted during the dilation stroke. The only way to have an engine in these conditions is to expend an even smaller amount of work in the compression stroke. This is only possible if the populations of levels $\pm2h$ are exponentially close to each other before the stroke, which in turn requires an exponentially large temperature $T_{a}$.

On the other hand, if $J < h_{b}/4 $, the exponential in Eq. (\ref{eq:upperE-asymp2}) goes quickly to zero for small $T_{b}$, and the RHS of this expression becomes  $< h_{a}$. However, this contradicts our starting hypothesis that $T_{a}\gg h_{a}$, indicating that there is in fact no physical solution. Indeed, as already noted in the main text (and seen in Figs. \ref{Fig-Phases}(a,b)), in this case the $E$-zone has no boundary at all for small $T_{b}$ (an engine always becomes possible as $T_{b}\rightarrow 0$, for any $T_{a}$).

\subsubsection{Boundary of the lower $E$-zone}
For the `lower'  $E$-zone ($T_{b} > T_{a}$) ,  we again look for solutions to  $f_{a} = f_{b}$, but in the limits where (i) $T_{a}$ is `small' and $T_{b}$ is `large' or (ii) $T_{a}$ and  $T_{b}$  are both `small'.\medskip

(i) For $T_{a} \ll 8J,  2h_{b}, 2h_{a} \ll T_{b}$ we obtain an equation analogous to Eq. (\ref{eq:upperEleft}) but with $a,b$ exchanged:
\begin{align}
\frac{h_{b}}{T_{b}}\left(1-\frac{2J}{T_{b}}\right) & =  \left(1+\exp\left(\frac{8J - 2h_{a}}{T_{a}}\right)\right)^{-1} 
\end{align}
Solving for $T_{a}$ this time, using the same approximation, 
we obtain 
\begin{equation}\label{eq:lowerE-asymp1}
T_{a} = \frac{8J-2h_{a}}{\ln(T_{b}+2J-h_{b})-\ln(h_{b})} %
\end{equation}
Consistently with Result 1, we see that a solution for an engine in this regime requires $J > h_{a}/4$. 

Note that Eq. (\ref{eq:lowerE-asymp1}) is a (slowly) \emph{decreasing} function of $T_{b}$. As discussed in the main text, this implies the very counterintuitive property that, for any fixed $T_{a}$, there is a finite value of $T_{b} > T_{a}$ above which the cycle ceases to function as an engine. \medskip

(ii) For $T_{a}, T_{b} \ll 2h_{b}, 2h_{a}  < 8J$, we use Eq. (\ref{eq:expapprox}) for both work functions, obtaining
\begin{align}\label{eq:rEcond}
\frac{8J - 2h_{a}}{T_{a}} = \frac{8J - 2h_{b}}{T_{b}}
\end{align}
or
\begin{align}\label{eq:lowerE-asymp2}
T_{a} = r_{E}T_{b}
\end{align}
with
\begin{align}\label{eq:re}
 r_{E} = \frac{4J - h_{a}}{4J- h_{b}}.
 \end{align}
Note  $0  < r_{E} <1$ for $J > h_{a}$.

\subsubsection{Boundaries of the $H$- , $R$- and $A$ -zones}

\begin{table}[t]
\begin{center}
\begin{tabular}{|c|c|c|c|}
 & $W_{cycle}$ & $Q_{hot}$ & $Q_{cold}$\\
Engine & $>0$ & $ >  0$ &$  < 0$ \\
Accelerator & $<0$ & $ > 0$ & $ < 0$ \\
Refrigerator & $<0$ &  $ <  0$ & $ >  0$ \\
Heater & $<0$ & $ <  0$ & $ <  0$ 
\end{tabular}
\end{center}
\label{default}
\caption{The four operational regimes allowed by the First and Second Laws for a thermal machine operating between two thermal reservoirs \cite{Buffoni19}. Note we use opposite sign conventions for heat and work.}
\end{table}%

It is worth recalling the conditions defining each zone, which are summarized in Table I above. In terms of energy exchanges, the difference between a \emph{heater} and an \emph{accelerator} is in the sign of  $Q_{hot}$. The boundary between the $H$- and $A$- zones is therefore defined by the condition $Q_{hot} = 0$. Similarly, the difference between a heater and a refrigerator is in the sign of $Q_{cold}$, so the boundary between the $H$- and $R$- zones is defined by the condition $Q_{cold} = 0$. 
For $T_{a}> T_{b}$, $Q_{hot} = Q_{a}$ and $Q_{cold} = Q_{b}$.  From Eq.  (\ref{eq:work1}), 
 these borders are thus described respectively by the equations
\begin{align}
8J (p_{1}^{b} - p_{1}^{a}) = 2h_{a} (f_{a} - f_{b}) \\
8J (p_{1}^{b} - p_{1}^{a}) = 2h_{b} (f_{a} - f_{b}) 
\end{align}

For  $2h_{j}/T_{j} \sim 8J/T_{j} \sim \varepsilon \ll 1$, we can use again an iterative method, as with Eq. (\ref{eq:upperE-asymp1}) above. The solutions to these equations have the linear forms
\begin{align}
T_{a}^{AH}= r_{AH} T_{b} +s_{AH} + O(\varepsilon) \label{eq:AH-asymp} \\
T_{a}^{HR} = r_{HR} T_{b} +s_{HR} + O(\varepsilon)
\label{eq:HR-asymp} 
\end{align}
where the angular coefficients are respectively
\begin{align}
r_{AH} = \frac{6J^{2} +h_{a}^{2}}{6J^{2} + h_{a}h_{b}} \label{eq:rah} \\
r_{HR} =  \frac{6J^{2} + h_{a}h_{b}} {6J^{2} +h_{b}^{2}} \label{eq:rhr}\end{align}
while the linear coefficients are respectively
\begin{align}
s_{AH} = -J\frac{12J^{2}(r_{AH}^{2}-1) - r_{HR}^{2}(h_{a}(2h_{a}+h_{b})+3h_{a}^{2}}{6J^{2}+h_{a}^{2}}\\
s_{HR} =-J \frac{12J^{2}(r_{HR}^{2}-1) + h_{a}(2h_{b}+h_{a})-3h_{b}^{2}r_{HR}^{2}
}{6J^{2}+h_{a}h_{b}}
\end{align}
Note that, for $T_{b} > T_{a}$, the roles of $Q_{a}, Q_{b}$ invert, so the solutions for the two boundaries are merely swapped ($T_{a}^{AH}$ would now have the expression in Eq. (\ref{eq:HR-asymp}), and vice-versa). However, since $r_{AH}$ and $r_{HR}$ are both $>1$, these solutions are incompatible with the initial hypothesis ($T_{b} > T_{a}$). This implies that, for large $T_{a} < T_{b}$, there can be no refrigerator or heater.

Finally, the difference between an \emph{accelerator} and a \emph{refrigerator} is of a different nature. In both cases one bath loses heat while the other one gains; the difference is whether the bath losing heat is the hotter or the colder one. Thus, the boundary between these zones is not marked by a change of sign in either bath's heat exchange, but in the difference of their temperatures. In other words, in this case the boundary is the line 
\begin{equation}\label{eq:AR}
T_{a}^{AR}  = T_{b},
\end{equation}
 irrespective of any asymptotic limits. Note this holds both for refrigerators rotating in the `ordinary' sense (clockwise, $T_{a} > T_{b}$), and in the `counter-rotating' sense (anti-clockwise, $T_{b} > T_{a}$).

It is easy to verify that
\begin{align}
r > r_{AH} > r_{HR } > 1. 
\end{align}
This shows that, for large enough $T_{b}$, if we start from $T_{a} = T_{b}$ and increase $T_{a}$, we must obtain in succession a refrigerator, a heater, an accelerator and finally an engine - which is indeed what we observe in Fig. \ref{Fig-Asympt}. In addition, it also shows that, for large enough $T_{b}$, a heater or refrigerator are only possible if $T_{a} > T_{b}$ (equivalently, for large enough $T_{b}> T_{a}$  only an accelerator or an engine are possible).

\subsubsection{Strong coupling limit}

If $J$ is large compared to $h_{a}, h_{b}$, then it is clear from 
Eqs. (\ref{eq:re}), (\ref{eq:rah}), (\ref{eq:rhr}) that $r_{E}$, $r_{AH}$ and $r_{HR}$ all tend to 1. As a consequence (i) both the $H$- and $R$- zones will collapse into an extremely narrow range around the line $T_{a} = T_{b}$ (property (vi) in section xxx). In addition: (ii) The lower $E$-zone border will also accompany this line for low $T_{b}$, and as a consequence this zone will increase in size. These effects can be seen in Fig. \ref{Fig-Phases}(d).

\subsection{Asymptotic Efficiency}

Another application of  Eqs. (\ref{eq:f-expansion})-(\ref{eq:expapprox2}) is to study the efficiency $\eta$ in the asymptotic limits where the bath temperatures $T_{a}$ and/or $T_{b}$ become very large or very small.

As we have shown in Fig. \ref{Fig-Eff-Limit}(a), as $T_{a}\rightarrow \infty$ (i,e., as $\beta_{a}\rightarrow 0$), $\eta$ tends to a value above $\eta_{0}$ (dashed line) only when $T_{b}$ is under a certain threshold.  To see why: suppose
\begin{align}
T_{b} \ll 8J < 2h_{b} < 2h_{a} \ll T_{a}.
\end{align}
In this case, we obtain, up to an error of order $\varepsilon$,
\begin{align}
W_{cycle} & = \frac{2(h_{a}-h_{b})\exp\left[\frac{2h_{b}-8J}{T_{b}}\right]}
           {1 + \exp\left[\frac{2h_{b}-8J}{T_{b}}\right]}\\
Q_{a} & =  \frac{8J+ 2 h_{a}\exp\left[\frac{2h_{b}-8J}{T_{b}}\right]}
           {1 + \exp\left[\frac{2h_{b}-8J}{T_{b}}\right]} -2J,
\end{align}
so 
\begin{align}  \label{eq:eta1}
 \eta  &=  \frac{2(h_{a}-h_{b})\exp\left[\frac{2h_{b}-8J}{T_{b}}\right]}
  {6J + 2(h_{a} - J) \exp\left[\frac{2h_{b}-8J}{T_{b}} \right]} 
 \end{align}

Choose a fixed value $J = \frac{h_{b}}{4}(1-\delta)$ for some $\delta < 1$. 
Substituting above, we obtain 
\begin{align}  
 \eta  &=  \frac{2(h_{a}-h_{b})e^{\frac{2h_{b}\delta}{T_{b}}}}
  {\frac{3}{2}h_{b}(1-\delta) + \big(2h_{a} - \frac{1}{2}h_{b}(1-\delta)\big) e^{\frac{2h_{b}\delta}{T_{b}}}} \nonumber\\
  \nonumber \\
         &=  \frac{\eta_{0}\,e^{\frac{2h_{b}\delta}{T_{b}}}}
  {\frac{3}{4r}(1-\delta) + \big(1 - \frac{1}{4r}(1-\delta)\big) e^{\frac{2h_{b}\delta}{T_{b}}}} \label{eq:eta2}
 \end{align}
 It follows that $\eta > \eta_{0}$ if and only if
 \begin{align}
  \frac{3}{4r}(1-\delta) <  \frac{1}{4r}(1-\delta) e^{\frac{2h_{b}\delta}{T_{b}}}
 \end{align}
 or
  \begin{align} \label{eq:TbThresh}
  T_{b} < \frac{2h_{b} \delta}{\ln 3}
  \end{align}
 (for consistency with the hypothesis that $T_{b} \ll 2h_{b}$, we see that we must in fact  choose $\delta~\ll~1$).
 
In other words, given a sufficiently high coupling strength, there exists a finite threshold for $T_{b}$, under which the engine efficiency converges to a higher than classical value in the limit of high $T_{a}$.  This can be seen in Fig. \ref{Fig-Eff-Limit}(a) above. For the parameters in that figure $(J = 0.24, h_{a} = 2, h_{b} = 1)$, it is easy to check that $\delta = 0.04$ and the threshold value is $T_{b} \sim 0.073$, which matches well with the exact curves. It is also worth noting that, in this limit, $W_{cycle}$ is not negligible; indeed $W_{cycle} \rightarrow 2(h_{a}-h_{b})$ for $T_{b} \ll 2 h_{b} \delta$.

Moreover, it can also be noted from the figure that, whenever $T_{b}$ is below the threshold, the efficiency actually \emph{increases further} with $\beta_{a}$ (ie, as  $T_{a}$ \emph{decreases}).  Let us now demonstrate that  this does indeed happen as $T_{a}$ is reduced from infinity. For this, is useful to look at 
 $\lim_{T_{a}\rightarrow \infty }\frac{\partial \eta}{ \partial \beta_{a}}$,  where $\beta_{a}  \equiv 1/T_{a}$ is the inverse temperature. We wish to show that this limit is strictly positive for $T_{b}$ satisfying Eq. (\ref{eq:TbThresh}). (Directly  calculating $\lim_{T_{a}\rightarrow \infty }\frac{\partial \eta}{ \partial T_{a}}$ does not give the same information, since in this limit $\frac{\partial \eta}{ \partial T_{a}} = - \frac{1}{T_{a}^{2}} \frac{\partial \eta}{ \partial \beta_{a}}  \rightarrow 0$).

Going back then to the exact equations for $Q_{a}$ and $W_{cycle}$ (Eq. (\ref{eq:work1})), replacing $J = \frac{h_{b}}{4}(1-\delta)$, and taking the appropriate limits, it is possible to show after some algebra that
\begin{align}
\lim_{T_{a}\rightarrow \infty} \frac{\partial \eta}{\partial \beta_{a}} = \frac{\eta}{W_{cycle}} \left[\eta\big(2h_{a}^{2} +\frac{3}{4} h_{b}^{2}(1-\delta)^{2} \big) - 2h_{a}(h_{a}-h_{b})  \right]
\end{align}

Using the asymptotic expression for $\eta$ in Eq. (\ref{eq:eta2}), we find that the term in square brackets is positive if
\begin{align}
 \exp\left[\frac{2 h_{b}\delta}{T_{b}}\right] > \frac{3}{1 + \frac{3}{2r}(1-\delta)}.
\end{align}
In other words: 
\begin{align}\label{eq:TbThresh2}
\lim_{T_{a}\rightarrow \infty} \frac{\partial \eta}{\partial \beta_{a}} > 0 \iff T_{b} < \frac{2h_{b}\delta}{\ln 3 - \ln \big(1 + \frac{3}{2r}(1-\delta) \big)}.
\end{align}
Since this upper bound is strictly larger than the one in Eq. (\ref{eq:TbThresh}), we arrive at our desired result.

For the parameters in Fig. \ref{Fig-Eff-Limit}(a), the threshold in Eq. (\ref{eq:TbThresh2}) equals approximately 0.144. It can be seen that $\eta$ does indeed initially increase with $\beta_{a}$ for all values of $T_{b}$ below this value  (including those under the smaller threshold in Eq. (\ref{eq:TbThresh})).

Of course, for sufficiently low $T_{a}$ an engine becomes impossible, so the efficiency must go to zero (at best, this must happen for $T_{a} < r T_{b}$). This implies that the efficiency must reach a maximum at some finite $T_{a}$, which is what we observe in the figure. 

In summary, we have proven\medskip

\textbf{Result 3}:  
Given a sufficiently high coupling strength $J = \frac{h_{b}}{4}(1-\delta)$, with $\delta \ll 1$, and a sufficiently low cold bath temperature $T_{b} \ll 8J$, and taking the limit $T_{a}\rightarrow \infty$, then the engine efficiency $\eta$ converges to a value higher than $\eta_{0}$ if and only if $T_{b}$ is under the threshold given by Eq. (\ref{eq:TbThresh}). In addition, in these circumstances $\eta$ increases further as $T_{a}$ is reduced from infinity, reaching a maximum at some finite $T_{a}$.

\medskip

\section{Extensions of the `idle level' scenario}

In the Otto engine scenario we discuss, levels can be divided into two groups: they are either `idle', independent of the adiabatic parameter $\lambda$, or else vary with $\lambda$ according to Eq.\,(\ref{eq:workinglevels}). This situation, which we will now refer to as the `idle scenario', is intended as a minimal departure from the uniform scenario' studied in Refs. \cite{Kosloff17, Geva92, Kieu04, Kosloff02}, where all levels obey this second property (i.e., all gaps scale proportionally to $\lambda$). In those cases, the efficiency always equals the standard value $\eta_{0}  = 1 - \lambda_{b}/\lambda_{a}$.  As we have shown, the presence of even a single `idle' level allows a host of interesting effects to appear, including efficiency beyond $\eta_{0}$, counter-rotating engines, a temperature gap where no engine is possible, etc. 

It by no means follows, however, that the idle scenario is the only one where such phenomena can occur, or that the presence of idle levels is crucial for them in any fundamental sense.
In this section, we seek to emphasise this point by briefly sketching three extensions of the idle scenario, none of which feature idle levels, but where many of the phenomena we have pointed out will continue to occur. This is not intended as a thorough exploration of these scenarios, let alone an exhaustive list of possibilities, but as an illustration and perhaps a starting point for further studies.

\subsection{Extension by Continuity}

A first, simple, possibility is to slightly deform the idle scenario. For example, suppose we introduce a small adiabatic shift to each (originally) idle level, so that they cease to be idle. Physically, this means these levels are now very weakly coupled to the work source/sink responsible for the time-dependence during the adiabatic strokes. The shifts do not all have to be equal, nor to apply to the `working' levels.

Now, by continuity, for sufficiently small deformations most properties of the original engine must still remain true. For instance, if $\eta > \eta_{0}$  before the deformation, or if the output work is $W > 0$ in a counter-rotating cycle, then these properties will continue to hold for a slightly deformed engine - even though the latter will no longer have true idle levels, nor the same energy gaps. How large these deformations can be made will of course depend on details of each specific Hamiltonian and adiabatic shifts. An analogous argument can also be made with respect to scenarios with imperfect thermalization in the isochoric strokes. 

\subsection{Extension by Symmetry}

It seems to be a widely known (but, to our knowledge, unpublished) fact that idealised Otto engines of the kind we discuss, i.e., with full thermalization and perfect adiabatic strokes, possess the following symmetry\footnote{This interesting result was pointed out to us by an unnamed referee of an earlier version of this manuscript, whom we thank.}: the heat exchanges with each bath, and hence the total output work, are invariant under \emph{independent} global energy shifts at either end of the adiabatic strokes. More precisely: \medskip

\textbf{Theorem:}  Consider an ideal Otto engine $E$, satisfying Eqs.\,(\ref{eq:Qgen}, \ref{eq:Wgen}), with energy levels $\{E_{n}^{a}\}$ at $\lambda = \lambda_{a}$ and $\{E_{n}^{b}\}$  at  $\lambda = \lambda_{b}$, respectively. Consider also another Otto engine $E'$, identical except for the energy shifts $E_{n}'^{a} = E_{n}^{a} + \delta_{a}$, $E_{n}'^{b} = E_{n}^{b} + \delta_{b}$, $\forall n$, where in general $ \delta_{a} \neq  \delta_{b}$.  Then $Q_{a} = Q_{a}', Q_{b} = Q_{b}', W_{cycle} = W'_{cycle}.$ \medskip

\emph{Proof:} The case $\delta_{a} = \delta_{b}$ is trivial, since it corresponds to a global energy shift of the entire scenario. For similar reasons, we can always, without loss of generality, choose $\delta_{b} = 0$, i.e., analyze only the effect of $\delta_{a}\neq 0$ (we assume this from now on).  Note now that $p_{n}'^{j} = p_{n}^{j}, \forall n,j$.  The result then follows straightforwardly from Eq.\,(\ref{eq:Qgen}), and from the fact that  $W'_{cycle} = Q'_{a} + Q'_{b} \;\;\square$ \smallskip

This symmetry implies, in particular, that engines $E$ and $E'$ will both behave exactly the same with respect to efficiency, regimes of operation, sense of rotation of the cycle, and more generally all the phenomena reflected in Fig. \ref{Fig-Phases}.

Nevertheless, a global shift in $E_{n}^{a}$  is not a complete symmetry of the engine. It does affect the separate work exchanges $W_{a\rightarrow b}$ and $W_{b\rightarrow a}$  in each adiabatic stroke (even though their sum is invariant). For instance
\begin{align*}
W'_{a\rightarrow b} & =   \sum_{n} p_{r
n}'^{a} \left(E_{n}'^{b} - E_{n}'^{a}\right)  \\
  & =   \sum_{n} p_{n}^{a} \left(E_{n}^{b} - E_{n}^{a} - \delta_{a}\right) \\
& =  W_{a\rightarrow b} - \delta_{a}
\end{align*} 
This shows the original and `shifted' engines describe \emph{physically different} situations, i.e., they are indeed different engines with similar properties. 

Note now finally that if engine $E$ operates in the `idle scenario', $E'$ will generally \emph{not}: for nearly all values of $\delta_{a}$, it will in fact have no idle levels whatsoever.

\subsection{Extension by Linearity}

Finally,  consider an engine whose levels can be divided into two groups (A and B), each of which shifts adiabatically according to Eq.\,(\ref{eq:workinglevels}), but with different linear coefficients, as follows:
\begin{align*}
&\text{Group A: }\;E_{n,A} (\lambda)= c_{n,A} \lambda \\
&\text{Group B: }\;E_{n,B} (\lambda)= \delta + c_{n,B} \lambda\;,
\end{align*}
where all $c_{n,j} \neq 0$ and where we can choose the constant $\delta >0$  without loss of generality.  Thus, for small enough $\lambda$, Group $B$ levels are all higher than those in Group $A$.

There are no idle levels in this scenario either. Nevertheless we can apply much of the same reasoning from that case. For example: following a derivation analogous to the one leading to Eq.\,(\ref{eq:enhancedeff}), we now obtain that, for $T_{a} > T_{b}$ and $Q_{a} >0$, an engine will have efficiency
\begin{align}
 \frac{\eta}{\eta_{0}}  = 1 - \delta \frac{ \sum_{n,B} \Delta p_{n,B}} {Q_{a}}. \label{eq:linearscenarioeff}
\end{align}
where $\Delta p_{n,B} =  p^a_{n,B} - p^b_{n,B}$ .  Once again, an increase with respect to the standard Otto efficiency $\eta_{0}$ is in principle possible, with no presence of energy-basis coherence etc. The condition for this increase to occur in this new scenario is however different, namely:  
\begin{align}
 \sum_{n,B} \Delta p_{n,B} < 0. 
\end{align} 
In other words, this engine will have a greater-than-standard efficiency only if the overall population of the $B$ levels decreases as the temperature increases from $T_{b}$ to $T_{a}$, and  $\lambda$ increases from $\lambda_{b}$ to $\lambda_{a}$. In particular, suppose, for simplicity, that the engine operates between values of $ \lambda$ that are sufficiently small that no level crossings occur between Group $A$ and Group $B$ levels. Then we need the population of the higher ($B$) levels to decrease when the temperature increases. 

Achieving this is not entirely obvious - for example, for a fixed Hamiltonian, the highest level of a finite-dimensional quantum system always \emph{gains} population with increasing temperature.
However the reverse is not impossible when the variation in $\lambda$ is also taken into account. For instance, if the highest levels have large positive $c_{n,B}$ (and thus shift significantly upward as $\lambda$ increases from $\lambda_{b}$ to $\lambda_{a}$), then their populations at $T_{a} > T_{b}$ may well become smaller than at $T_{b}$, despite the increase in temperature. A simple 3-level example where this indeed happens (with, e.g., the highest level alone constituting `Group B', and the other two `Group A') is as follows: 
\begin{align*}
E_{1,A} (\lambda)&=  \lambda \\
E_{2,A} (\lambda)&= 2 \lambda \\
E_{1,B} (\lambda)&= 1 + 4 \lambda 
\end{align*}
It can be easily checked that, for example, if $T_{b}=1,T_{a}=4, \lambda_{b} = 1,  \lambda_{a} = 2$, then $p_{1B}$ decreases from about $1.3\%$ to about $1.1\%$ as $\lambda$ and $T$ both increase from $b$ to $a$.

\newpage

\end{document}